\documentclass[fleqn,usenatbib]{mnras}

\usepackage{newtxtext,newtxmath}

\usepackage[T1]{fontenc}

\DeclareRobustCommand{\VAN}[3]{#2}
\let\VANthebibliography\thebibliography
\def\thebibliography{\DeclareRobustCommand{\VAN}[3]{##3}\VANthebibliography}

\usepackage{graphicx}	

\usepackage{amsmath}	
\usepackage{amssymb}	
\usepackage[normalem]{ulem}
\usepackage{booktabs} 

\newcommand{\orcid}[1]{\href{https://orcid.org/#1}{\includegraphics[width=10pt]{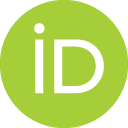}}}

\title[\texttt{zeus}: Lightning Fast MCMC]{\texttt{zeus}: A Python implementation of Ensemble Slice Sampling for efficient Bayesian parameter inference}

\author[M. Karamanis et al.]{
Minas Karamanis \orcid{0000-0001-9489-4612},$^{1}$\thanks{E-mail: minas.karamanis@ed.ac.uk}
Florian Beutler \orcid{0000-0003-0467-5438},$^{1}$
and John A. Peacock \orcid{0000-0002-1168-8299}$^{1}$
\\
$^{1}$Institute for Astronomy, University of Edinburgh, Royal Observatory, Blackford Hill, Edinburgh EH9 3HJ, UK
}

\date{Accepted XXX. Received YYY; in original form ZZZ}

\pubyear{2021}

\begin{document}
\label{firstpage}
\pagerange{\pageref{firstpage}--\pageref{lastpage}}
\maketitle

\begin{abstract}
We introduce \texttt{zeus}, a well-tested \texttt{Python} implementation of the Ensemble Slice Sampling (ESS) method for Bayesian parameter inference. ESS is a novel Markov chain Monte Carlo (MCMC) algorithm specifically designed to tackle the computational challenges posed by modern astronomical and cosmological analyses. In particular, the method requires only minimal hand--tuning of $1-2$ hyper-parameters that are often trivial to set; its performance is insensitive to linear correlations and it can scale up to 1000s of CPUs without any extra effort. Furthermore, its locally adaptive nature allows to sample efficiently even when strong non-linear correlations are present. Lastly, the method achieves a high performance even in strongly multimodal distributions in high dimensions. Compared to \texttt{emcee}, a popular MCMC sampler, \texttt{zeus} performs $9$ and $29$ times better in a cosmological and an exoplanet application respectively.
\end{abstract}




\section{Introduction}
\label{sec:intro}

Over the past few decades the volume of astronomical and cosmological data has increased substantially. In response to that, a variety of astrophysical models have been developed to explain the plethora of observations. Markov chain Monte Carlo (MCMC) has been established as the standard procedure of inferring the model parameters subject to the available data in a Bayesian framework. Within the Bayesian context, the object that quantifies the probability distribution of the model parameters $\theta$ given the data $D$ and model $\mathcal{M}$ is the posterior distribution $\mathcal{P}(\theta)\equiv P(\theta | D, \mathcal{M})$ which is defined using Bayes's theorem:
\begin{equation}
    \label{eq:bayes}
    \mathcal{P}(\theta) = \frac{\mathcal{L}(\theta) \pi (\theta)}{\mathcal{Z}},
\end{equation}
where $\mathcal{L}(\theta) \equiv P(D|\theta, \mathcal{M})$ is the likelihood function, $\pi (\theta) \equiv P (\theta | \mathcal{M})$ is the prior distribution of the model parameters $\theta$, and $\mathcal{Z}\equiv P(D|\mathcal{M})$ is the, so called, Bayesian model evidence or marginal likelihood and in this context can be treated as a simple normalisation constant.

MCMC does not in general require knowing the value of the model evidence and it only depends on the ability to evaluate the unnormalised posterior distribution for arbitrary values of $\theta$. MCMC methods can then be used to generate (Markov) chains of samples from the posterior distribution. Those samples can be used to calculate integrals (e.g. parameter uncertainties, marginal distributions etc.) that are paramount for modern astronomical and cosmological analyses.

The most commonly used MCMC methods are variants of the Metropolis-Hastings (MH) algorithm \citep{metropolis1953equation, hastings1970}. MH consists of two steps. First, given the last sample in the chain, a new sample is proposed and then the Metropolis criterion determines whether or not that new sample should be accepted and thus added to the chain. The resulting chain is Markovian in the sense that each sample is proposed based only on the previous sample. The purpose of the Metropolis acceptance criterion is to bias the chain so that the time spent in a region of the parameter space would be proportional to the posterior probability in that region. In other words, the stationary distribution of the Markov chain is the target distribution i.e. the posterior distribution. For a detailed introduction to MCMC methods we direct the reader to \citet{mackay2003information} and for an intuitive introduction to Bayesian inference to \citet{jaynes2003probability}.

Arguably, the most difficult part of the MH algorithm is the proposal step. There are many ways of choosing a new sample and the efficiency of the method depends on this choice. By far the simplest one is the use of a normal (Gaussian) distribution, centred around the previous sample to generate the new proposed sample. The resulting method is often called Random Walk Metropolis algorithm and its performance is highly sensitive to the $n(n+1)/2$ elements that form its covariance matrix. Those elements generally need to be chosen \textit{a priori} or be adaptively tuned. More efficient methods utilise the gradient of the target distribution \citep{2017arXiv170102434B} or an ensemble of parallel and communicating chains \citep{gilks1994adaptive,ter2006markov, ter2008differential, goodman2010ensemble}.

Out of the methods mentioned in the previous paragraph we will focus our attention on the last one, the ensemble or population MCMC variety. The reason is simple: the Random Walk Metropolis algorithm requires a great amount of tuning (or \textit{a priori} knowledge) for it to perform efficiently and even then there is no guarantee that the proposal covariance matrix is optimal for the whole parameter space. On the other hand, gradient based methods, although very powerful, are in general unsuitable for astronomical applications in which the models that are used are almost always not differentiable.

One benefit of ensemble MCMC over its alternatives is that the ensemble of parallel chains (also known as walkers) collectively sample the posterior, thus information about their distribution can be shared and used to make better educated proposals. Other advantages include the lack of hand-tuning of hyper-parameters and their capacity for parallel implementation. For the aforementioned reasons, ensemble MCMC methods have dominated astronomical analyses. The most common ones are affine--invariant ensemble sampling (AIES) \citep{goodman2010ensemble} and differential evolution MCMC (DEMC) \citep{ter2006markov, ter2008differential}, both implemented in the popular \texttt{Python} package \texttt{emcee} \citep{foreman2013emcee,foreman2019emcee}.

In this paper we introduce \texttt{zeus}, a stable and well-tested \texttt{Python} implementation of Ensemble Slice Sampling (ESS) \citep{karamanis2020ensemble}. ESS is a method based on the ensemble MCMC paradigm, with the crucial difference being that its proposals are performed via Slice Sampling updates \citep{neal2003slice} instead of Metropolis-Hastings ones. As we will thoroughly demonstrate in Section \ref{sec:tests}, this subtle difference leads to substantial improvements in terms of sampling efficiency and robustness. \texttt{zeus} is a user-friendly tool that does not require any hand-tuning or preliminary runs and can scale up to 1000s of CPUs without any extra effort from the user.

\texttt{zeus} has been used in various astronomical and cosmological analyses, including cosmological tests of gravity \citep{Tamosiunas2020}, relativistic effects and primordial non-Gaussianity \citep{Wang2020}, 21cm intensity mapping \citep{Umeh2021}, and has been implemented as part of the \texttt{CosmoSIS} package \citep{Zuntz2015}.

\texttt{zeus} is open source software that is publicly available at \url{https://github.com/minaskar/zeus} under the \texttt{GPL-3 Licence}. Detailed documentation and examples on how to get started are available at \url{https://zeus-mcmc.readthedocs.io}.

\section{Ensemble Slice Sampling}
\label{sec:ess}

\texttt{zeus} is a \texttt{Python} implementation of the Ensemble Slice Sampling (ESS) method presented in \citet{karamanis2020ensemble}. 
Here we will provide a high-level description of the method and will refer to the accompanying paper for more details about the underlying algorithmic structure and mathematics.

ESS combines the ensemble MCMC paradigm with slice sampling. Since the use of slice sampling in astronomical parameter inference is rare we will start by explaining its function and how it differs from Metropolis updates. Then we will move on to discuss how it can be efficiently combined with ensemble MCMC.

\subsection{Slice sampling}

Slice sampling is based on the idea that sampling from a distribution with density $P(x)$ is equivalent to uniform sampling from the area under the plot of $f(x)\propto P(x)$. To this end, we introduce an auxiliary variable $y$, called height, such that the joint distribution $P(x,y)$ is uniform over the region $U=\lbrace (x,y): 0<y<f(x) \rbrace$. To sample from the marginal distribution $P(x)$, we first sample from $P(x,y)$ and then we marginalise by dropping the $y$ value of each sample.

\begin{figure}
    \centering
	\centerline{\includegraphics[scale=0.9]{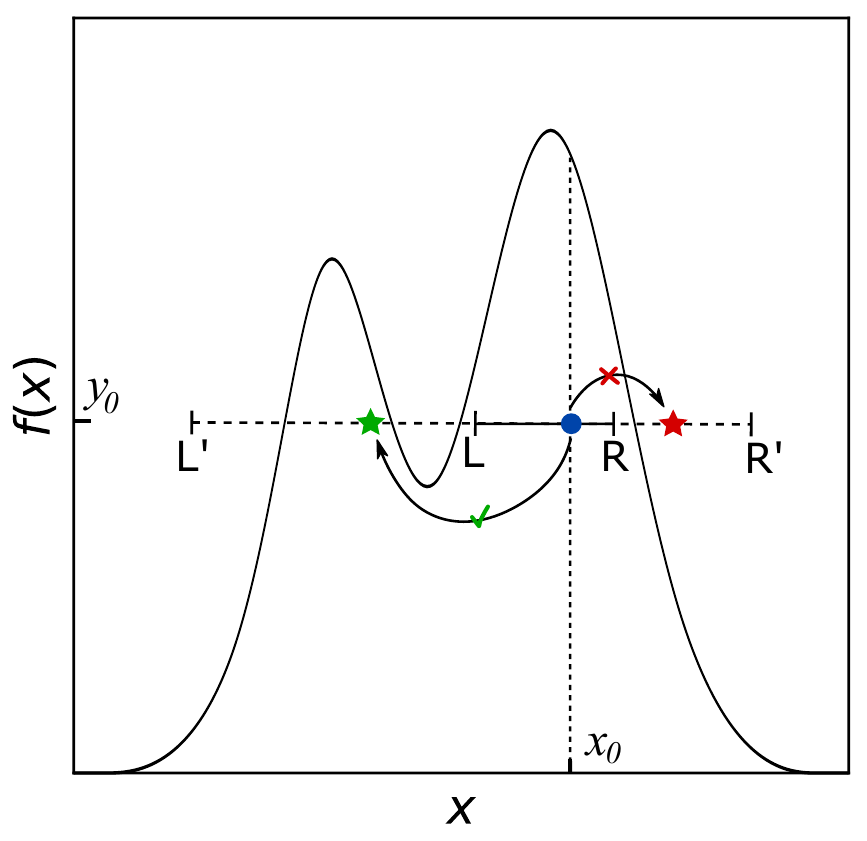}}
    \caption{Illustration of the univariate slice sampling update. Given the current sample $x_{0}$, a value $y_{0}$ is uniformly sampled along the vertical slice $(0,f(x_{0}))$ (dashed line) thus defining the initial point (blue). An interval $(L,R)$ is uniformly positioned horizontally around $(x_{0},y_{0})$ and it is expanded in steps of size $R-L$ until both its ends are outside the slice. The new sample is generated by  repeatedly sampling (uniformly) from the interval $(L',R')$ until a sample (green star) is found inside the slice. Samples outside of the slice (red star) are rejected and they are instead used to shrink $(L',R')$.}
    \label{fig:slice}
\end{figure}

In order to generate samples from $P(x,y)$ we utilise the following scheme \citep{neal2003slice}:
\begin{enumerate}
    \item Given the current state $x_{0}$, draw $y_{0}$ uniformly from $(0,f(x_{0}))$.
    \item Find an interval $I=(L,R)$ that contains all, or at least part, of the slice $s=\lbrace x: y_{0}<f(x)\rbrace$.
    \item Draw the new sample $x_{1}$ uniformly from $I\cap S$.
\end{enumerate}
To construct the interval $I$ (step ii), \citet{neal2003slice} introduced the stepping-out procedure that works by randomly positioning an interval of length $\mu$ around the sample $x_{0}$ (i.e. blue dot in Figure \ref{fig:slice}) and then expanding it in steps of size $\mu$ until both its ends (i.e. $L'$ and $R'$) are outside the slice. To obtain $x_{1}$ (i.e. green star in Figure \ref{fig:slice}) we then use the shrinking procedure in which candidates are sampled uniformly from $I$ until a point inside the slice $S$ is found. Samples outside of the slice are used to shrink the interval $I$. The two procedures are shown in Figure \ref{fig:slice}.

The length scale $\mu$ is the only free hyperparameter of slice sampling and although its choice can reduce or increase the computational cost of the method it generally does not affect its mixing properties (e.g. convergence rate, autocorrelation time, etc.). \texttt{zeus} utilises a stochastic optimization algorithm similar to \citet{tibbits2014automated} and based on the \citet{robbins1951stochastic} optimisation scheme in order to tune $\mu$ to its optimal value (see Section 3.1 of \citealt{karamanis2020ensemble} for more details).

It is important to note here that for multimodal target distributions there is no guarantee that the approximate slice would cross any of the other modes. In particular, if the initial estimate of the length scale $\mu$ is low then the probability of missing the other peaks, assuming that they are located far away, is also low. As we will show in Section \ref{sec:tests}, unlike simple slice sampling, ESS and thus \texttt{zeus} does not suffer from this effect.

\subsection{Walkers, moves and parallelism}

\begin{figure}
    \centering
	\centerline{\includegraphics[scale=0.9]{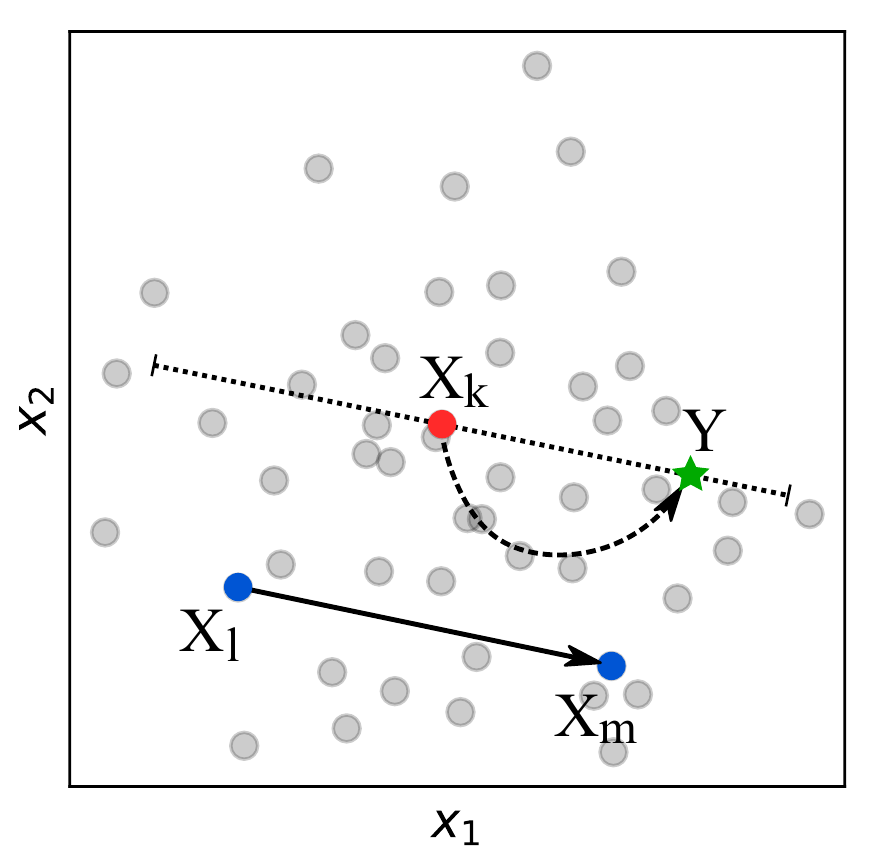}}
    \caption{The figure illustrates the differential move in the context of Ensemble Slice Sampling. The walker $X_{k}$ to be updated is shown in red. Two walkers, $X_{l}$ and $X_{m}$, (blue) are uniformly selected from the complementary ensemble (grey). The approximate slice (dotted line) is constructed parallel to the two walkers $X_{l}$ and $X_{m}$ using the stepping-out procedure. The new position $Y$ (green) of $X_{k}$ is sampled using the shrinking procedure along the approximate slice.}
    \label{fig:move}
\end{figure}

The slice sampling update described in the previous paragraphs is a univariate update scheme. For it to be used to sample from multivariate target distributions it needs to be generalised accordingly. Perhaps the simplest such generalisation in a multivariate setting is the use of slice sampling to sample along each coordinate axis in turn (i.e. component-wise slice sampling) or to sample along randomly selected directions in parameter space \citep{mackay2003information}. Although valid, both of these approaches are unsuitable in cases of correlated parameters in which the proper choice of direction can substantially accelerate mixing.

To address this issue, \citet{tibbits2014automated} proposed to orthogonalise the parameter space using the sample covariance, thus getting rid of linear correlations between parameters. We will instead follow a different, perhaps more flexible, approach to construct an efficient slice sampler. Our aim is to utilise an ensemble of parallel chains/walkers that can exchange information about the covariance structure of the target distribution and thus by-pass the difficulties posed by correlations.

As hinted in the introduction, the ensemble of walkers collectively sample the target distribution and thus their positions encode information about the correlations between the parameters. One way to take advantage of this information is to use it to construct direction vectors along which slice sampling can take place. Many moves that generate direction vectors from the complementary ensemble are possible. \texttt{zeus} offers a collection of them, including some that utilise clustering algorithms and density estimation methods. As we will show in Section \ref{sec:tests}, such moves can help accelerate sampling in difficult cases such as strongly multimodal distributions. Any distribution of the complementary ensemble can be used as a valid proposal to generate such direction vectors and \texttt{zeus} offers a highly flexible interface for the user to define such a move or choose one (or a mixture) from the ones that are already implemented and tested. Here is a list of the currently implemented moves in \texttt{zeus}:

\begin{itemize}
    \item \textbf{Differential move:} This is the default move used by \texttt{zeus} and shown in Figure \ref{fig:move}. Using the differential move, Ensemble Slice Sampling updates the position of each walker in the ensemble by slice sampling along a direction defined by the difference between two uniformly selected walkers from the rest of the ensemble (i.e. the complementary ensemble).
    \item \textbf{Gaussian move:} The Gaussian move samples the direction vectors along which slice sampling is performed from a normal distribution that shares the same covariance structure as the complementary ensemble. This approach is very efficient in cases in which the target distribution is close to normal.
    \item \textbf{Global move:} The Global move utilises a Dirichlet Process Gaussian Mixture to fit the complementary ensemble and proposes directions along different peaks of the target distribution in cases of strong multi-modality.
    \item \textbf{KDE move:} The KDE move samples the direction vectors from a Gaussian Kernel Density Estimate of the complementary ensemble. This can be useful in cases of highly non-Gaussian target distributions.
    \item \textbf{Random move:} The Random move performs slice sampling along isotropic directions. This is equivalent of standard multivariate slice sampling and it is mostly offered for testing purposes as it cannot handle correlations efficiently.
\end{itemize}
For more information on how those moves work as well as a comparison of the Differential, Gaussian and Global moves we direct the interested reader to \citet{karamanis2020ensemble}. Unless stated otherwise the Differential move will be used for the following examples.

To parallelise this process and capitalise on the availability of multiple CPUs we randomly split the ensemble into two sets of walkers (i.e. active and passive sets)~\citep{foreman2013emcee} and choose to update the positions of the active walkers along direction vectors defined by passive walkers. Then the passive become active and \textit{vice versa} and the process is repeated. The ensemble splitting technique is required in order to parallelise the algorithm without violating detailed balance. Parallelisation is achieved in practise using either \texttt{multiprocessing} or \texttt{MPI} using the implemented \texttt{ChainManager} utility that can distribute both multiple ensembles and multiple chains in parallel computing environments at the same time. Heuristics to determine the number of required walkers per application are discussed in Section \ref{sec:discussion}.

\section{Empirical Evaluation}
\label{sec:tests}

For the empirical evaluation of \texttt{zeus} we use five toy examples that manifest significant aspects of real astronomical applications\footnote{For additional demonstrations on similarly common structures (e.g. the funnel) we direct the reader to the accompanying paper \citep{karamanis2020ensemble}.} (i.e. linear and non-linear correlations, multimodality, heavy tails, hard boundaries) and two real-world astronomical examples characteristic of modern astronomical analyses.

\subsection{Toy examples}

\begin{figure*}
    \centering
	\centerline{\includegraphics[scale=0.55]{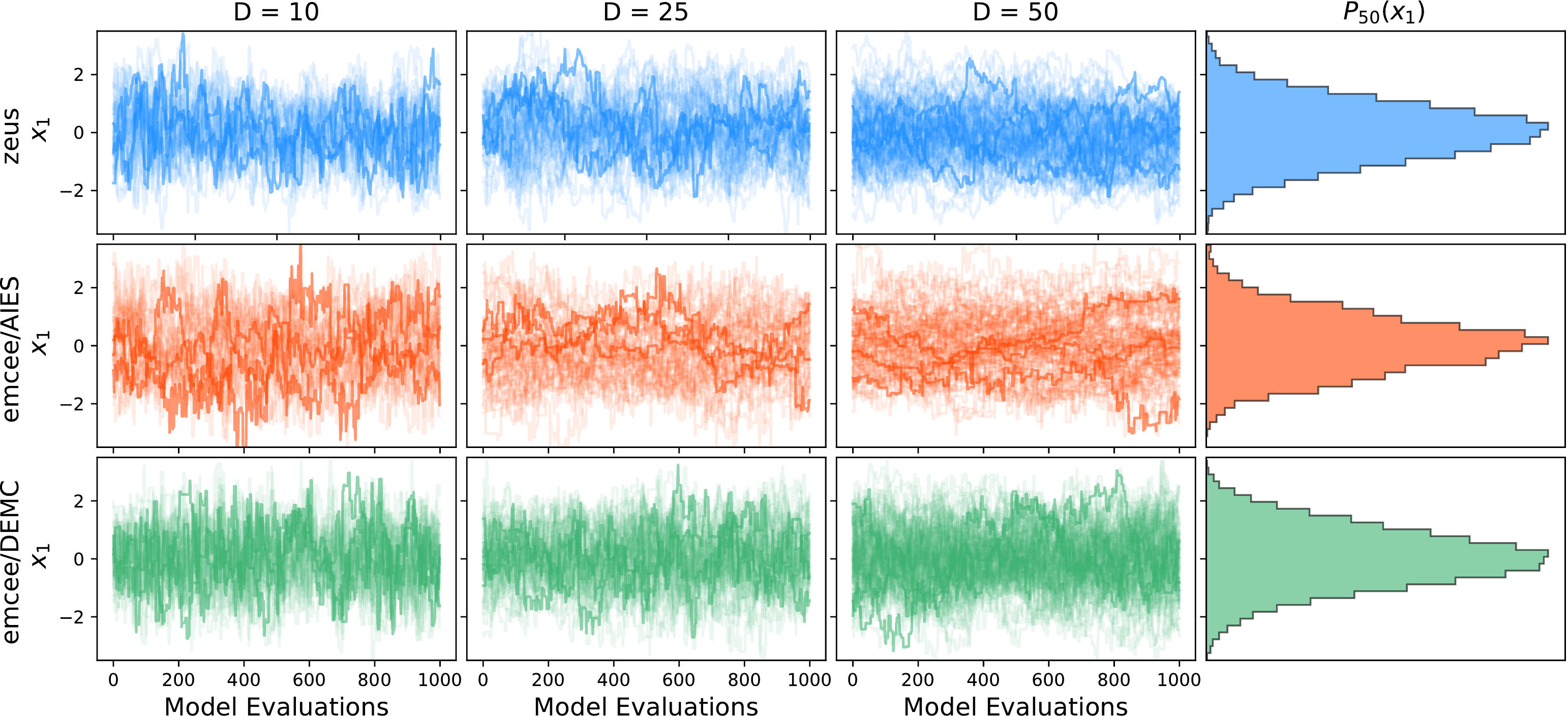}}
    \caption{The figure shows numerical results (i.e. walker trajectories/chains for the first parameter) demonstrating the performance of the three ensemble MCMC methods in the case of a normal (Gaussian) target distribution in $10, 25$ and $50$ dimensions respectively. The last column illustrates the 1-D marginal posterior corresponding to the first parameter $x_{1}$ estimated directly from the samples for the 50-dimensional case.}
    \label{fig:gaussian}
\end{figure*}
\begin{figure}
    \centering
	\centerline{\includegraphics[scale=0.47]{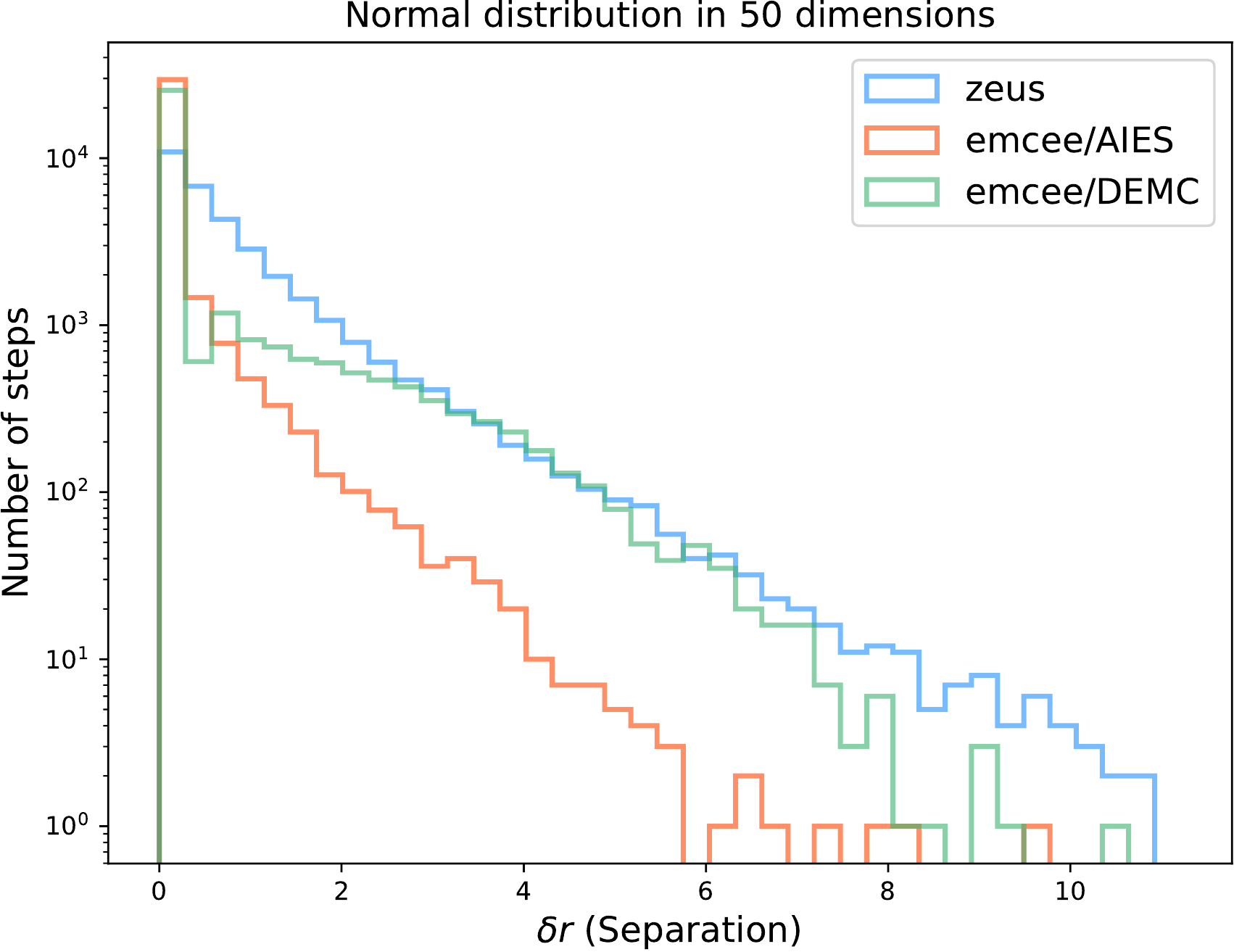}}
    \caption{This figure shows the distribution of step sizes of walkers for the three different samplers in the case of a normal (Gaussian) target distribution in $D=50$. It is important to note here that both \texttt{emcee} algorithms exhibit a peak at zero separation; \texttt{zeus} on the other hand does not due to its non-rejection nature.}
    \label{fig:gaussian_sep}
\end{figure}

\begin{figure*}
    \centering
	\centerline{\includegraphics[scale=0.55]{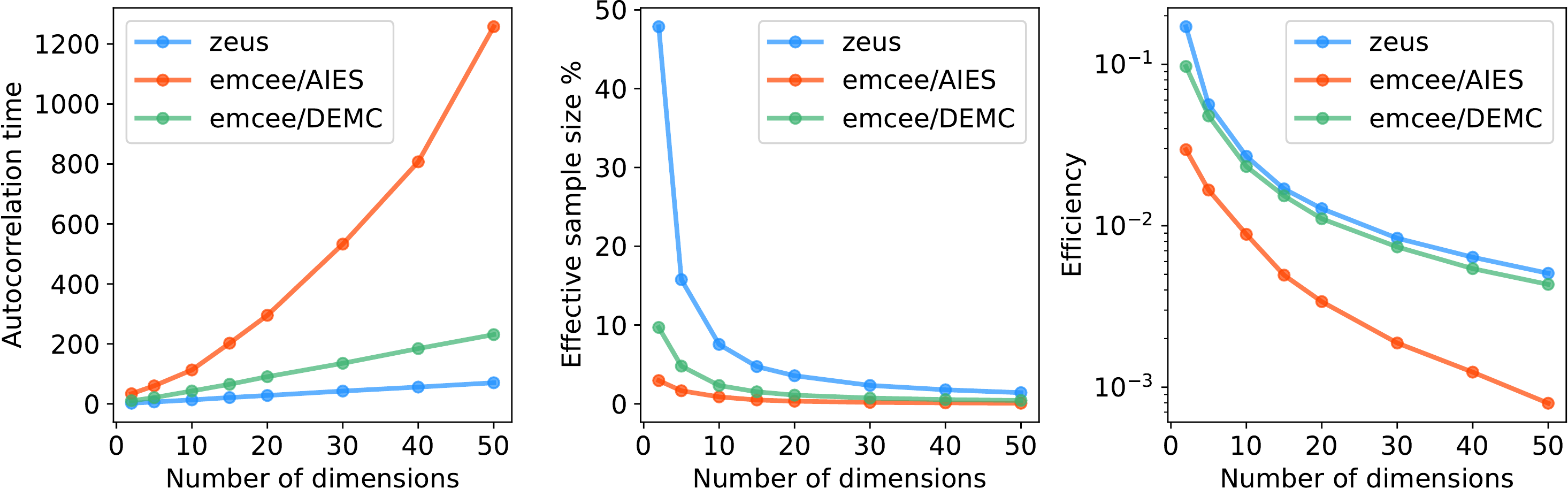}}
    \caption{The figure shows numerical estimates of the integrated autocorrelation time (number of steps along a chain required to obtain an independent sample; left panel), the effective sample size (percentage of effectively independent samples in a chain; middle panel), and the sampling efficiency (i.e. effective sample size per model evaluation; right panel) for a normal target distribution and varying number of dimensions. The number of walkers was set to $4\times D$ for \texttt{zeus} and $16\times D$ for \texttt{emcee}, this was the optimal choice (i.e. the one maximising the efficiency for the given dimensionality) for each sampler. \texttt{zeus} and \texttt{emcee/DEMC} exhibit linear scaling of the autocorrelation time with the number of dimensions whereas \texttt{emcee/AIES} scales exponentially.}
    \label{fig:act}
\end{figure*}

In order to understand the behaviour of \texttt{zeus} in various sampling scenarios it is important to study its performance in different toy examples that demonstrate different characteristics of common target distributions that arise in astronomical applications. For that reason we chose five such toy examples. The first one is a normal (Gaussian) distribution which by definition is characterised only by linear correlation between its parameters. The second toy problem is the ring distribution, a characteristic example of strong non-linear correlations. The third example is a Gaussian mixture with two components. While the purpose of the first two examples is to study the behaviour of the algorithm in the presence of linear and non-linear correlations respectively, the goal of the third example is to demonstrate the ability of \texttt{zeus} to sample efficiently from multimodal target distributions. The fourth toy example investigates the effect that heavy tails have on the sampling efficiency and the fifth the shows the effects that hard boundaries have on sampling.

We compare \texttt{zeus} with two popular alternatives offered by \texttt{emcee}, namely affine--invariant ensemble sampling with the \textit{stretch} move (\texttt{emcee}/AIES) and the differential evolution move (\texttt{emcee}/DEMC). The main goal of this analysis is to justify our choice of slice sampling as the basis of \texttt{zeus} instead of Metropolis updates through the use of simple yet instructive toy examples.

For all three toy examples discussed below we adopt the same analysis procedure, where we initialise the walkers by sampling from a normal distribution $\mathcal{N}(\mathbf{0}, \mathbf{I})$ where $\mathbf{I}$ is the identity covariance matrix and we discarded $10^4$ iterations as burn-in. 

The main metric that we use to investigate the behaviour of the samplers in those toy examples and to compare their performance is the distribution of steps performed by the walkers. As a step we define the distance spanned in parameter space by a single walker in a single iteration. This is a fundamental measure of the efficiency of an MCMC method and it is directly related to the expected squared jump distance (ESJD;~\citealt{pasarica2010adaptively}) given by:
\begin{equation}
    \label{eq:esjd}
    \text{ESJD} = \mathbf{E}\left[|\theta_{t+1}-\theta_{t}|^{2}\right]= 2 \,( 1 - \rho_{1}) \cdot\text{Var}_{(\pi)}(\theta_{t})\,,
\end{equation}
where $\theta_t$ are the chain samples, $\rho_{1}$ is the first-order autocorrelation, and $\text{Var}_{(\pi)}(\theta_{t})$ is a function of the stationary distribution only. Assuming that the higher-order autocorrelations $\rho_{2}, \rho_{3}, \dots$ are monotonically decreasing with respect to $\rho_{1}$, then maximising the ESJD leads to minimisation of the autocorrelation between chain elements and thus maximisation of the sampling efficiency. In other words, the further away (i.e. the greater the ESJD) the walkers jump per iteration,  the higher the sampling efficiency of the method. A benefit of using ESJD instead of the autocorrelation time as a metric is that the former, as an expectation value, is more accurate when computed using short chains.

\begin{figure}
    \centering
	\centerline{\includegraphics[scale=0.47]{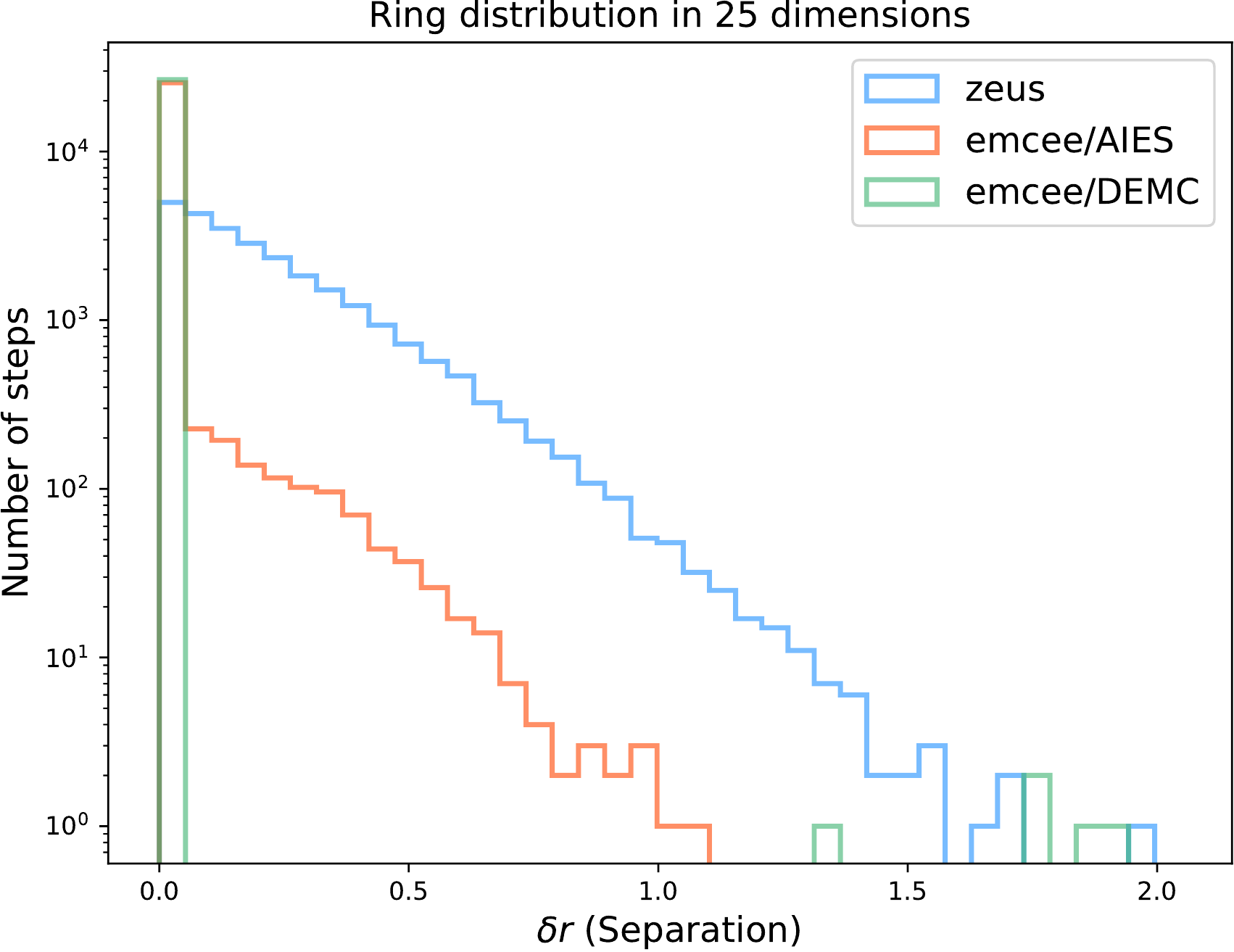}}
    \caption{This figure shows the distribution of step sizes of walkers for the three different samplers in the case of a ring target distribution in $D=25$. It is important to note here that both \texttt{emcee} algorithms exhibit a peak at zero separation; \texttt{zeus} on the other hand does not. The existence of the zero-peak in \texttt{emcee} is due to the high number of rejected proposals (i.e. low acceptance rate).}
    \label{fig:ring_sep}
\end{figure}

\begin{figure}
    \centering
	\centerline{\includegraphics[scale=0.47]{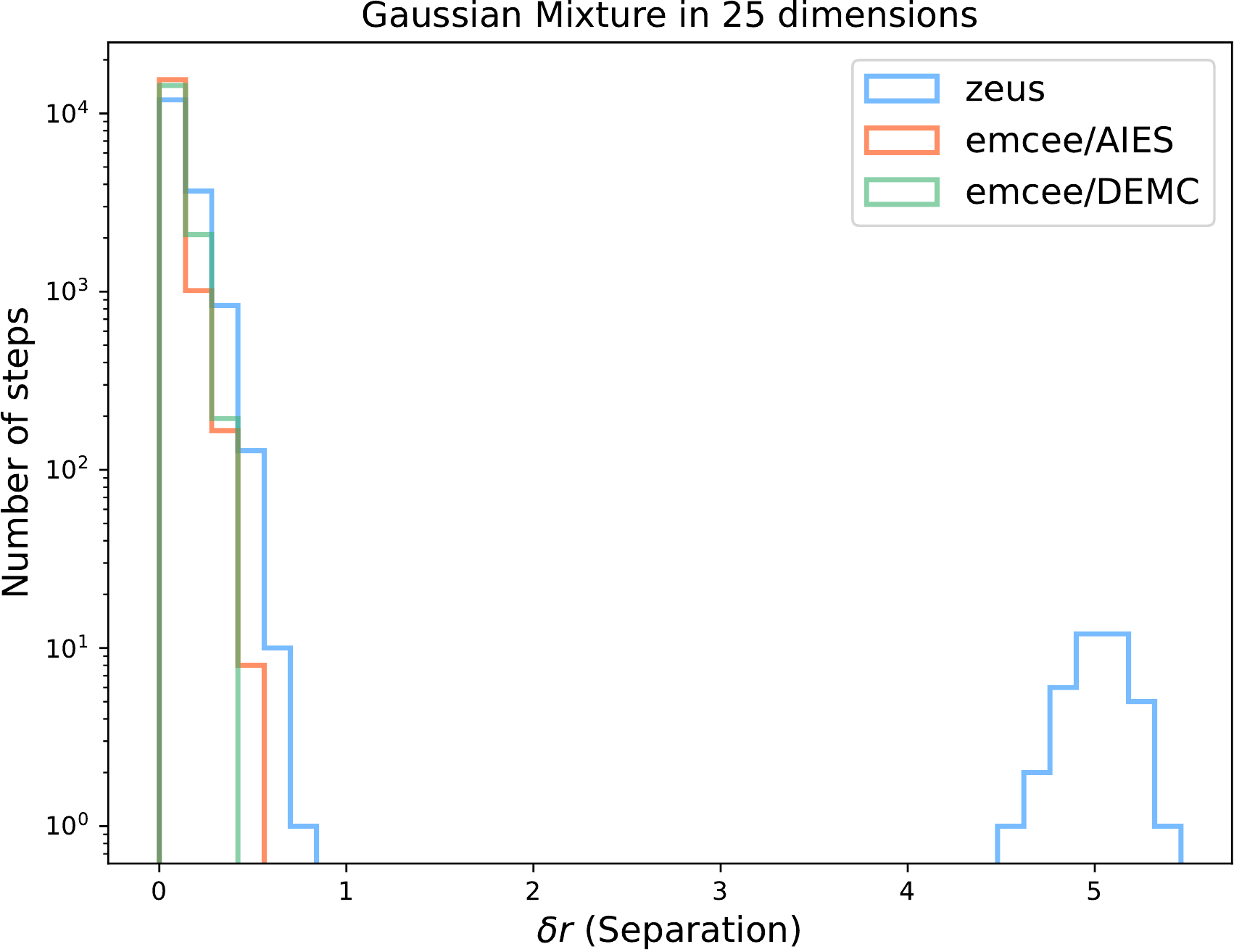}}
    \caption{This figure shows the distribution of step sizes of walkers for the three different samplers in the case of a two-component  Gaussian mixture target distribution in $D=25$. It is important to note here that both \texttt{emcee} algorithms exhibit a peak at zero separation; \texttt{zeus} on the other hand does not due to its non-rejection basis.}
    \label{fig:bimodal_sep}
\end{figure}

\begin{figure}
    \centering
	\centerline{\includegraphics[scale=0.47]{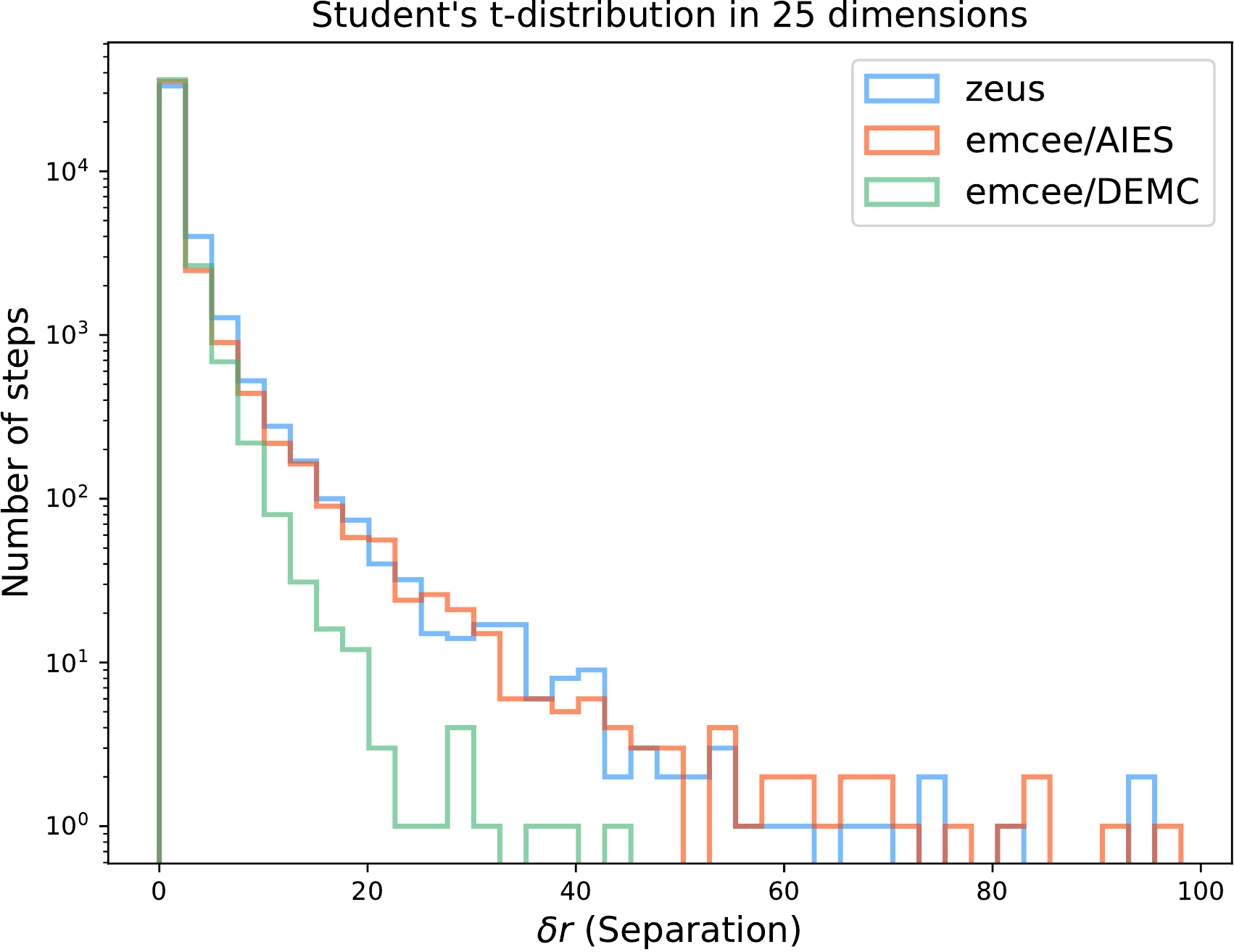}}
    \caption{This figure shows the distribution of step sizes of walkers for the three different samplers in the case of the Student's $t$-distribution with $2$ degrees of freedom in $D=25$. \texttt{zeus} and \texttt{emcee}/AIES exhibit similar distributions whereas \texttt{emcee}/DEMC performs shorter steps.}
    \label{fig:student_sep}
\end{figure}

\begin{figure}
    \centering
	\centerline{\includegraphics[scale=0.47]{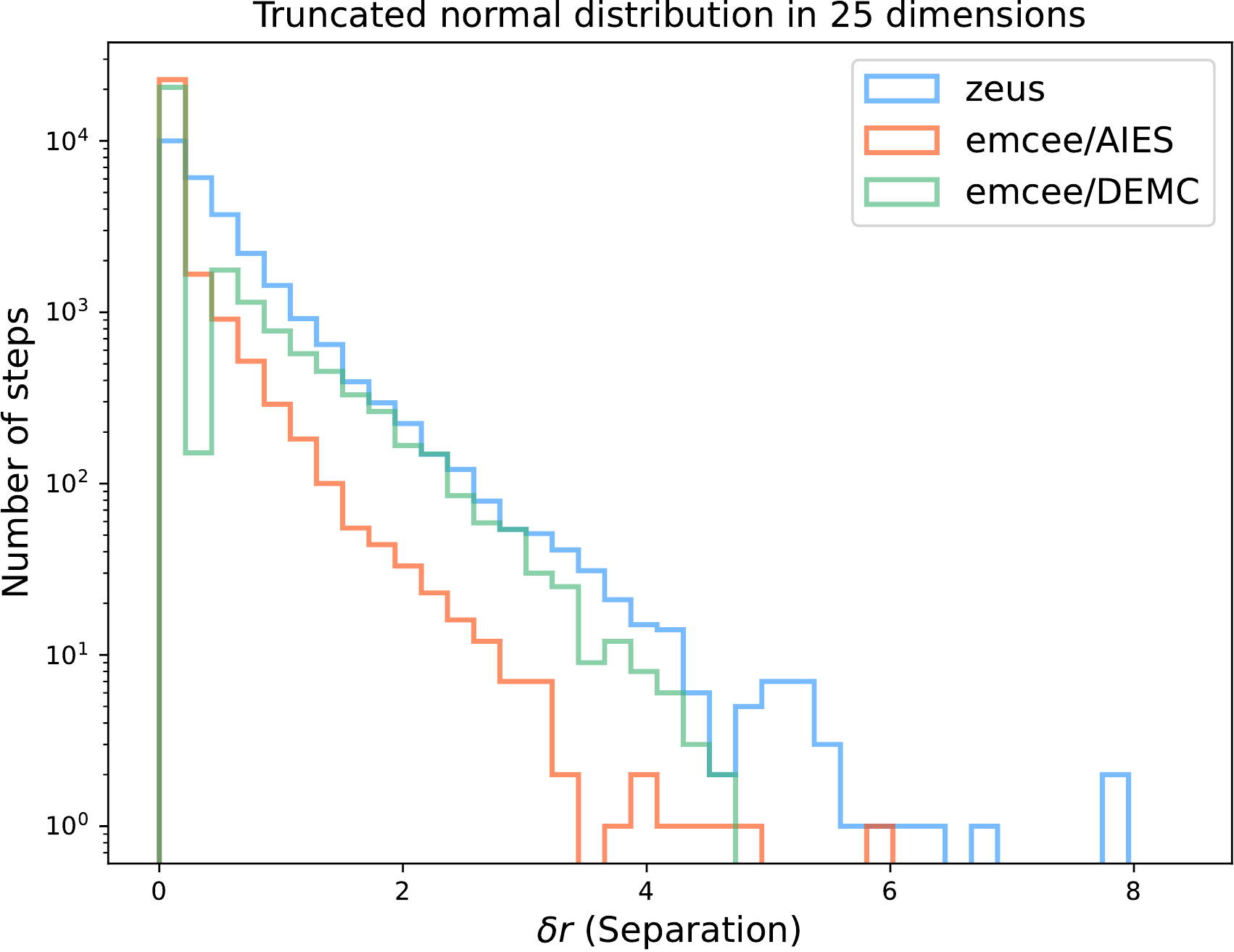}}
    \caption{This figure shows the distribution of step sizes of walkers for the three different samplers in the case of the truncated normal distribution in $D=25$. \texttt{zeus} and \texttt{emcee}/AIES exhibit similar distribution whereas \texttt{emcee}/DEMC performs shorter steps.}
    \label{fig:truncated_sep}
\end{figure}

In order to account for the different computational cost (i.e. different number of model evaluations per iteration) between \texttt{zeus} and \texttt{emcee} we thinned the chains of the latter method according to the average number of model evaluations of \texttt{zeus}. This allowed us to compare the distribution of steps of the three samplers as shown in Figures \ref{fig:gaussian_sep}, \ref{fig:ring_sep}, \ref{fig:bimodal_sep}, \ref{fig:student_sep}, and \ref{fig:truncated_sep} for the five toy examples respectively.

\subsubsection{The correlated normal distribution}

Starting with the normal target distribution it is important to note here that all three of the methods used in the comparison are affine--invariant\footnote{Differential evolution Metropolis is only approximately affine--invariant due to the jitter that it is often added to its proposal. This however has a negligible effect.}, meaning that their performance is immune to any linear correlations between the parameters. Since the normal distribution incorporates, by construction, only linear correlations (i.e. the 2D marginal distribution contours look like ellipses), it is the perfect testing ground to assess the effect that high dimensionality has on the three methods independently of other complications. For our example we used a zero-mean normal distribution with a covariance matrix in which the diagonal elements are set to $1$ and the off-diagonal ones are equal to $0.95$. We then proceed by sampling the aforementioned distribution in $10$, $25$ and $50$ dimensions. Based on Figure \ref{fig:gaussian} one can see that the walkers of \texttt{emcee}/AIES dissolve into an inefficient random walk characterised by low step size and high autocorrelation time as the number of parameters increases. \texttt{zeus} and \texttt{emcee}/DEMC are not so severely affected by the high number of parameters exhibiting a substantially lower autocorrelation.

Let us now try to explain this difference in behaviour by looking into the distribution of the steps of the walkers in Figure \ref{fig:gaussian_sep}. One thing to notice here is that the distribution of the steps of \texttt{zeus}'s walkers extends significantly further away than those of \texttt{emcee}/AIES and \texttt{emcee}/DEMC. This should come as no surprise since the construction of the approximate slice allows for larger steps than Metropolis updates as shown in Table \ref{tab:table1}. This is because when a proposal is rejected in slice sampling the approximate slice shrinks and another sample is proposed instead. In this way \texttt{zeus}'s walkers always move and the chance of staying fixed is zero -- unlike MH-based updates in which frequent rejection of samples is a necessity. This aforementioned procedure leads to greater steps in parameter space. The difference between \texttt{emcee}/AIES and \texttt{emcee}/DEMC is attributed to the fact that DEMC uses a proposal scale\footnote{The proposal scale $\gamma$ is similar to $\mu$ used in ESS in the sense that its value determines the length scale of the proposed jumps in parameter space. A high value would lead to large steps that are often rejected and a low value would lead to small steps that are often accepted but do not carry the walkers far. For such methods, a balance must me found.} $\gamma = 2.38 / \sqrt{D}$ that guarantees a constant acceptance rate accounting for the number of dimensions $D$. This proposal scale is however optimal only in the case of a normal target distribution such as the one that we are studying here and there is no guarantee that it would return acceptable results in non-Gaussian distributions. For the case of \texttt{emcee}/AIES, the relevant proposal scale $\gamma$ is allowed to vary in the range between $1/\alpha$ and $\alpha$ where $\alpha = 2$ is often taken as the typical value. It is clear that in the latter case $\gamma$ does not possess the desired scaling $\gamma\propto 1/\sqrt{D}$ and thus, although the method generates proposals in the right overall direction, most of the samples do not reside in the typical set \citep{speagle2019conceptual}. In other words, the lack of proper scaling of the proposal scale with the number of dimensions leads to \texttt{emcee}/AIES ``overshooting'' the typical set where most of the posterior mass is located.

We can also draw some useful insights about the sampling efficiency of those samplers and its scaling with the number of dimensions by estimating the integrated autocorrelation time of the chains. Given the autocorrelation time we can also estimate the effective sample size as the percentage of effectively independent samples in a chain. Dividing the effective sample size with the computational cost of each method we can then estimate the sampling efficiency. The results of such a comparison are shown in Figure \ref{fig:act}. We immediately notice here that the autocorrelation times of \texttt{zeus} and \texttt{emcee}/DEMC scale linearly with the number of dimensions, whereas the autocorrelation time of \texttt{emcee}/AIES scales exponentially. The computational cost of \texttt{zeus} per iteration per walker, although somewhat higher than that of \texttt{emcee}, does not vary with the number of dimensions. This means that in high dimensions, \texttt{zeus} dominates over \texttt{emcee}/AIES in terms of sampling efficiency.

\begin{figure*}
    \centering
	\centerline{\includegraphics[scale=0.55]{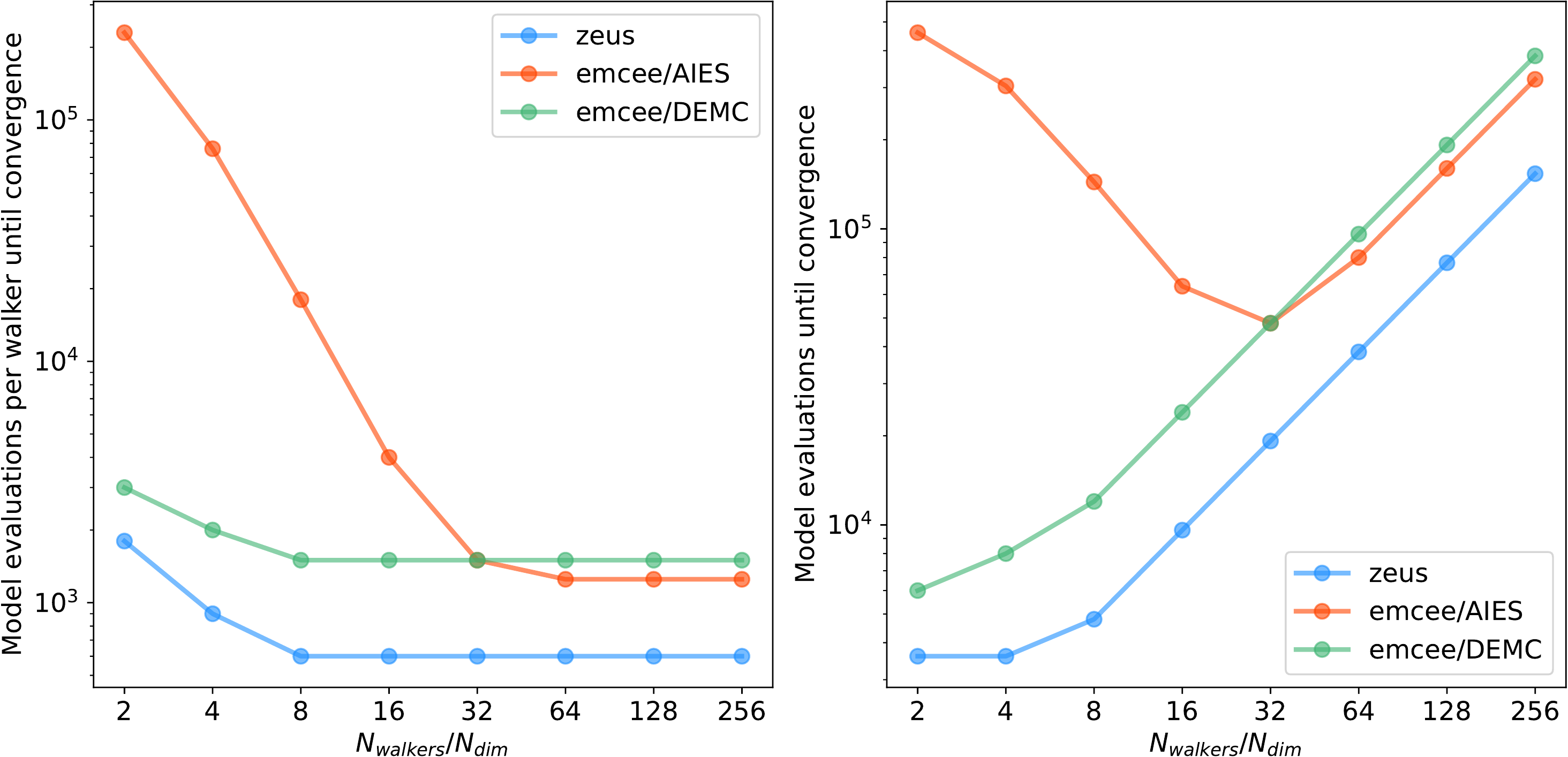}}
    \caption{The figure shows the computational cost until convergence is reached in terms of the number of model evaluations for the different ensemble samplers for a highly correlated 25--dimensional normal distribution. The left panel shows the computational cost for a single walker. From this we can see that the cost for a single walker decreases as we increase the number of walkers until it reaches a plateau. The high computational cost for low numbers of walkers can be attributed to the low variety or sparsity of possible proposals; this is significantly higher for \texttt{emcee}/AIES. The right panel takes into account the linear scaling of the total computational cost as we increase the number of walkers and shows the total computational cost for the whole ensemble until it converges.}
    \label{fig:convergence}
\end{figure*}
\begin{figure}
    \centering
	\centerline{\includegraphics[scale=0.47]{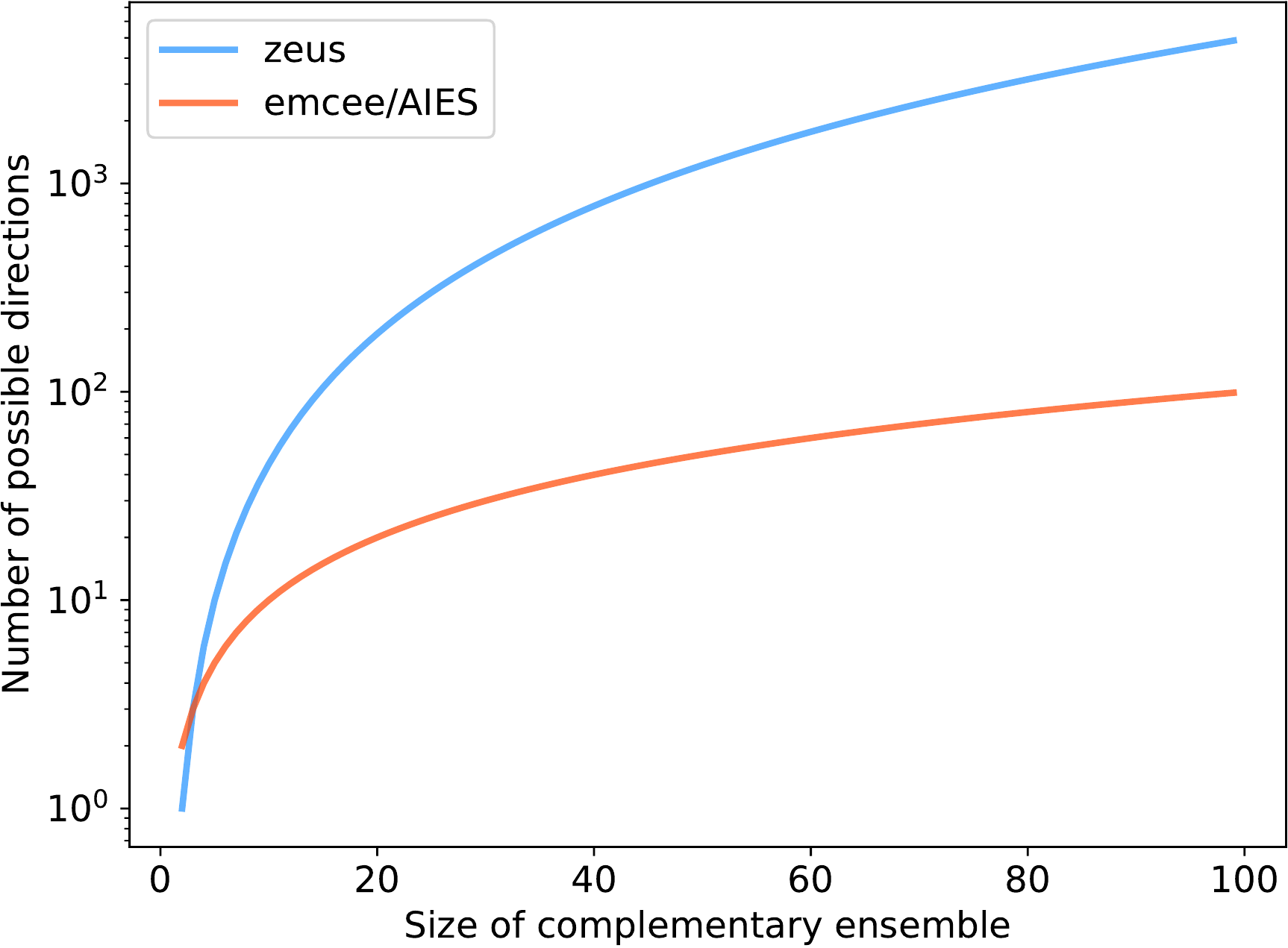}}
    \caption{The figure shows the number of possible directions along which \texttt{zeus} and \texttt{emcee}/AIES can propose new samples as a function of the number of walkers in the complementary ensemble. \texttt{emcee}/DEMC exhibits the same number of proposals as \texttt{zeus} and it is not plotted here. \texttt{zeus} has a much higher variety of possible directions compared to \texttt{emcee}/AIES for any given number of walkers, assuming that that number is greater than $2$.}
    \label{fig:proposals}
\end{figure}
\begin{table}
    \centering
    \caption{The table shows a comparison of \texttt{emcee}/AIES, \texttt{emcee}/DEMC and \texttt{zeus} in terms of the expected squared jump distance (ESJD; higher is better) for the five toy examples i.e. $50$-$D$ normal distribution, $25$-$D$ ring distribution, $25$-$D$ Gaussian mixture, $25$-$D$ Student's $t$-distribution, and $25$-$D$ truncated normal distribution.}
    \def\arraystretch{1.1}
    \begin{tabular}{lcccc}
        \toprule[0.75pt]
         & \texttt{emcee}/AIES   & \texttt{emcee}/DEMC   & \textbf{zeus}  \\
        \midrule[0.5pt]
        Normal    &    $0.5288$    &    $1.1162$    &   $\mathbf{2.1354}$   \\
        Ring   &    $0.0043$   &    $0.0006$    & $\mathbf{0.1257}$  \\
        Mixture   &    $0.0037$    &    $0.0056$    &  $\mathbf{0.1015}$ \\
        Student   &    $12.9124$    &    $2.4137$    &  $\mathbf{23.5720}$ \\
        Truncated   &    $0.0940$    &    $0.3501$    &  $\mathbf{0.5882}$ \\
        \bottomrule[0.75pt]
        \end{tabular}
    \label{tab:table1}
\end{table}
\begin{figure*}
    \centering
	\centerline{\includegraphics[scale=0.55]{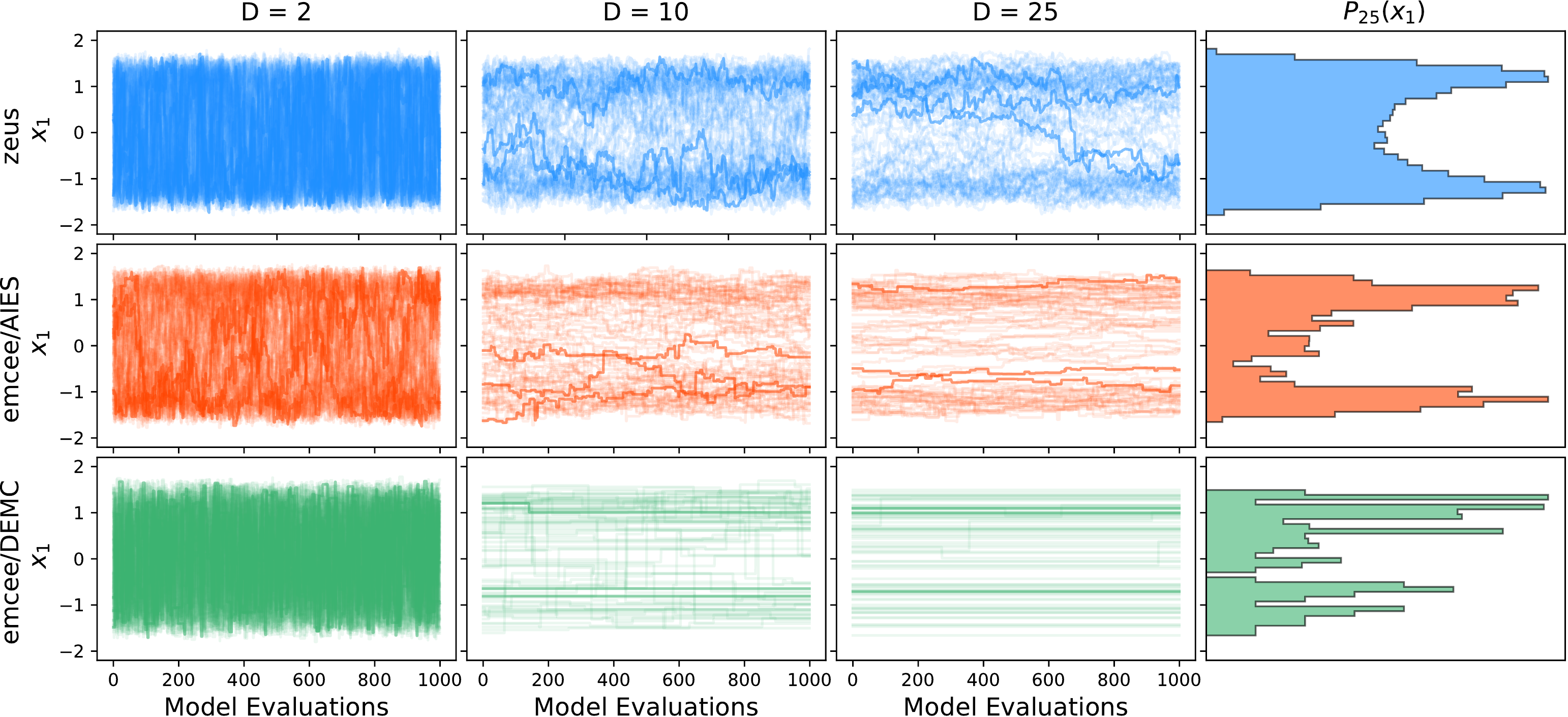}}
    \caption{The figure shows numerical results (i.e. walker trajectories/chains for the first parameter) demonstrating the performance of the three ensemble MCMC methods in the case of the ring target distribution in $2, 10$ and $25$ dimensions respectively. The last column illustrates the 1-D marginal posterior corresponding to the first parameter $x_{1}$ estimated directly from the samples for the 25-dimensional case. One can notice here that in $10$ and $25$ dimensions both \texttt{emcee} methods mix very slowly. In the 25-dimensional case almost all of \texttt{emcee}/DEMC's walkers are unable to move and the autocorrelation time is effectively infinite.}
    \label{fig:ring}
\end{figure*}

\begin{figure*}
    \centering
	\centerline{\includegraphics[scale=0.55]{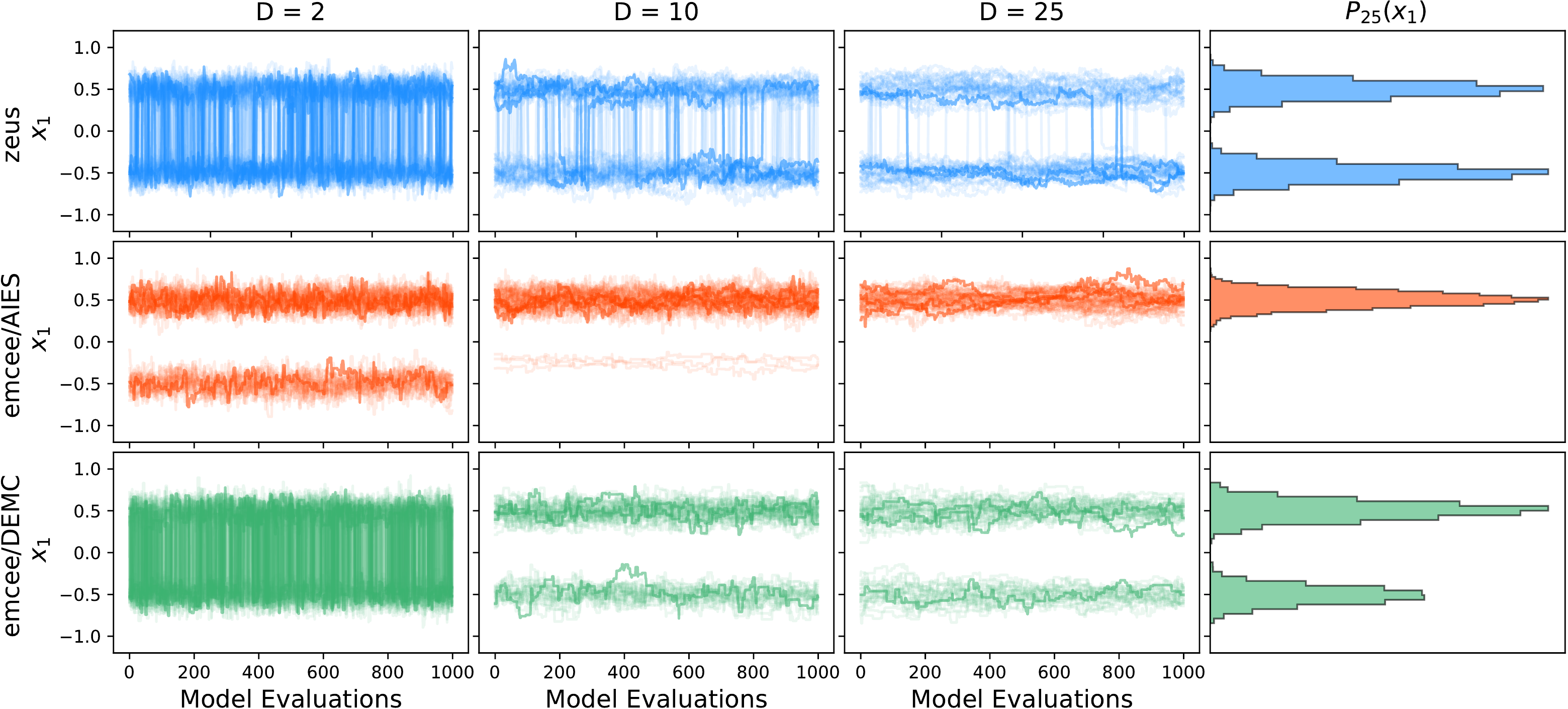}}
    \caption{The figure shows numerical results (i.e. walker trajectories/chains for the first parameter) demonstrating the performance of the three ensemble MCMC methods in the case of a two-component  Gaussian mixture target distribution in $2, 10$ and $25$ dimensions respectively. The last column illustrates the 1-D marginal posterior corresponding to the first parameter $x_{1}$ estimated directly from the samples for the 25-dimensional case. Whereas all three samplers make valid within-mode proposals, it is only \texttt{zeus} that manages to perform between-mode jumps and thus sample correctly from the target distribution in the 10 and 25-dimensional cases. Between-mode jumps are paramount in order to distribute the probability mass correctly between different modes.}
    \label{fig:bimodal}
\end{figure*}

\begin{figure*}
    \centering
	\centerline{\includegraphics[scale=0.55]{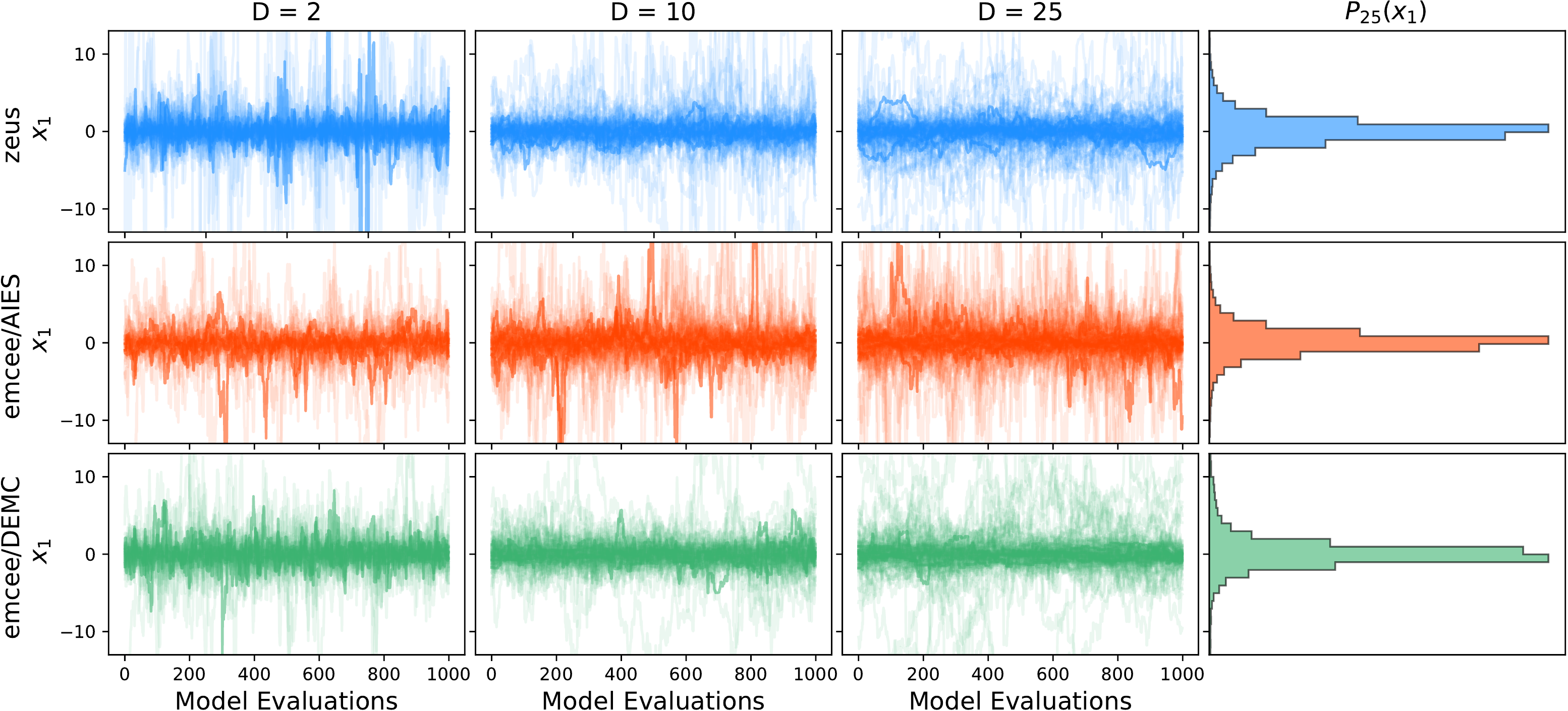}}
    \caption{The figure shows numerical results (i.e. walker trajectories/chains for the first parameter) demonstrating the performance of the three ensemble MCMC methods in the case of the Student's $t$-distribution with $2$ degrees of freedom in $2, 10$ and $25$ dimensions respectively. The last column illustrates the 1-D marginal posterior corresponding to the first parameter $x_{1}$ estimated directly from the samples for the 25-dimensional case.}
    \label{fig:student}
\end{figure*}

\begin{figure*}
    \centering
	\centerline{\includegraphics[scale=0.55]{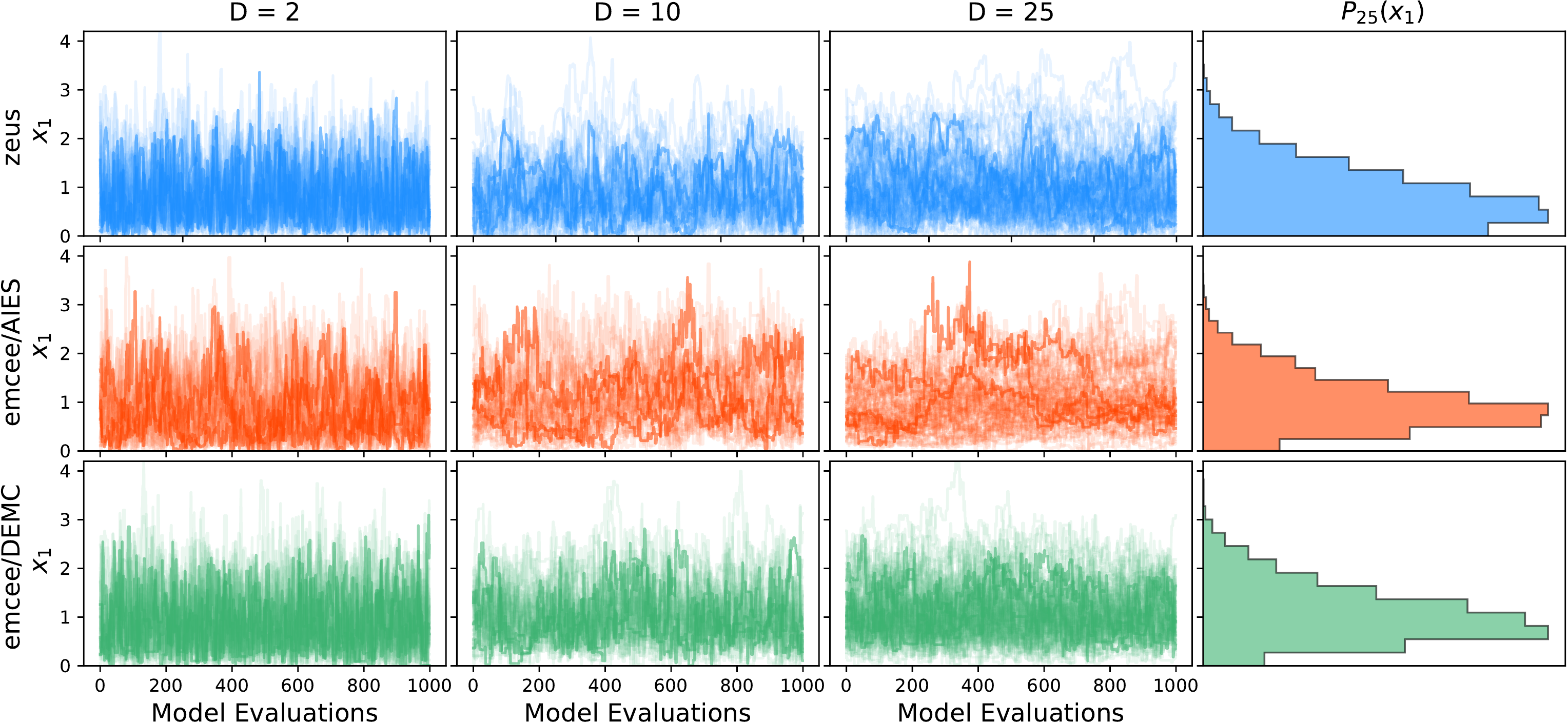}}
    \caption{The figure shows numerical results (i.e. walker trajectories/chains for the first parameter) demonstrating the performance of the three ensemble MCMC methods in the case of the truncated normal distribution in $2, 10$ and $25$ dimensions respectively. The last column illustrates the 1-D marginal posterior corresponding to the first parameter $x_{1}$ estimated directly from the samples for the 25-dimensional case. \texttt{zeus} exhibits the least amount of bias near the hard boundary at zero compared to \texttt{emcee}/AIES and \texttt{emcee}/DEMC.}
    \label{fig:truncated}
\end{figure*}

\begin{figure*}
    \centering
	\centerline{\includegraphics[scale=0.32]{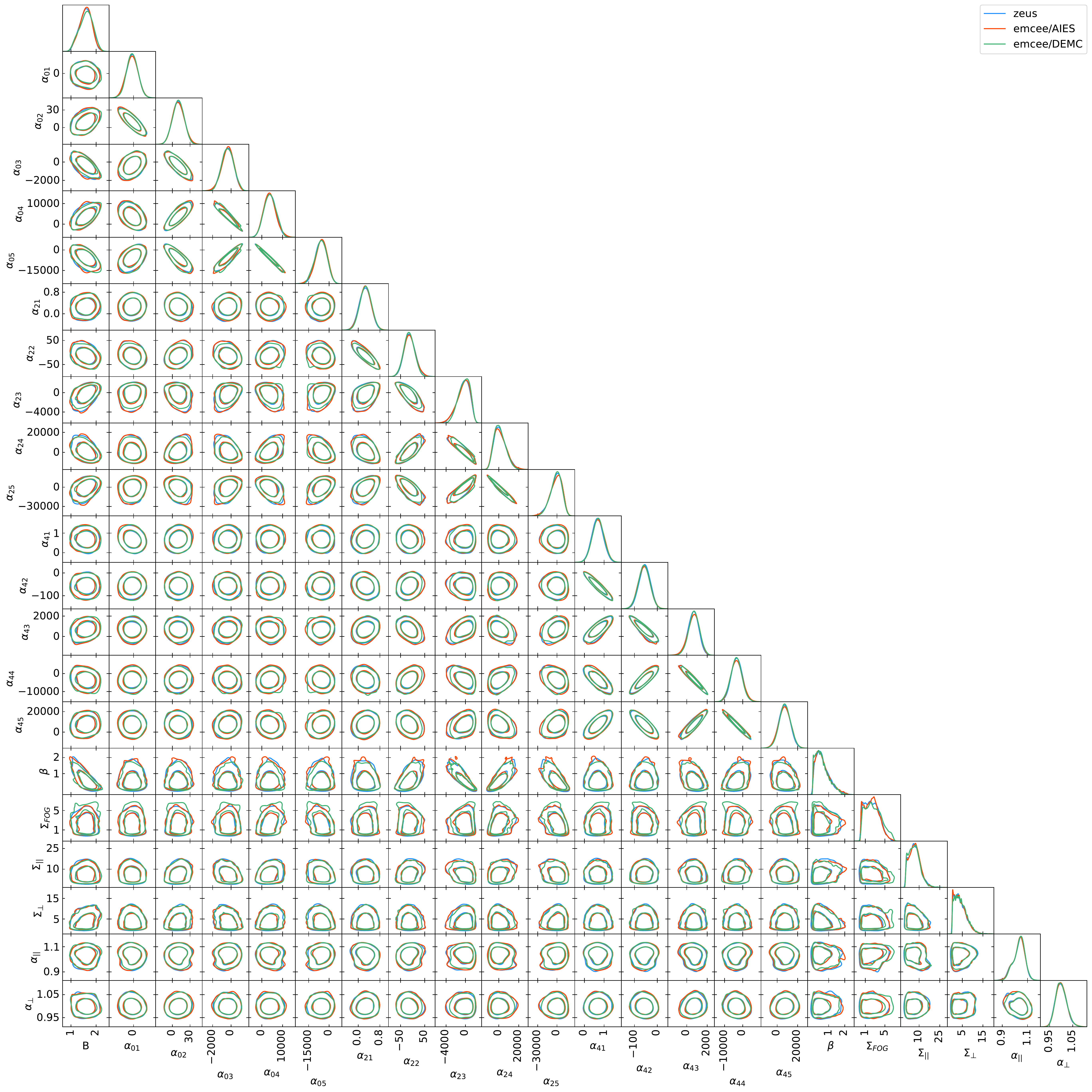}}
    \caption{A corner plot showing the 1-D and 2-D marginalised posteriors for the 22-parameter Baryon Acoustic Oscillation model as produced by the three different ensemble MCMC methods.}
    \label{fig:bao}
\end{figure*}

\begin{figure*}
    \centering
	\centerline{\includegraphics[scale=0.51]{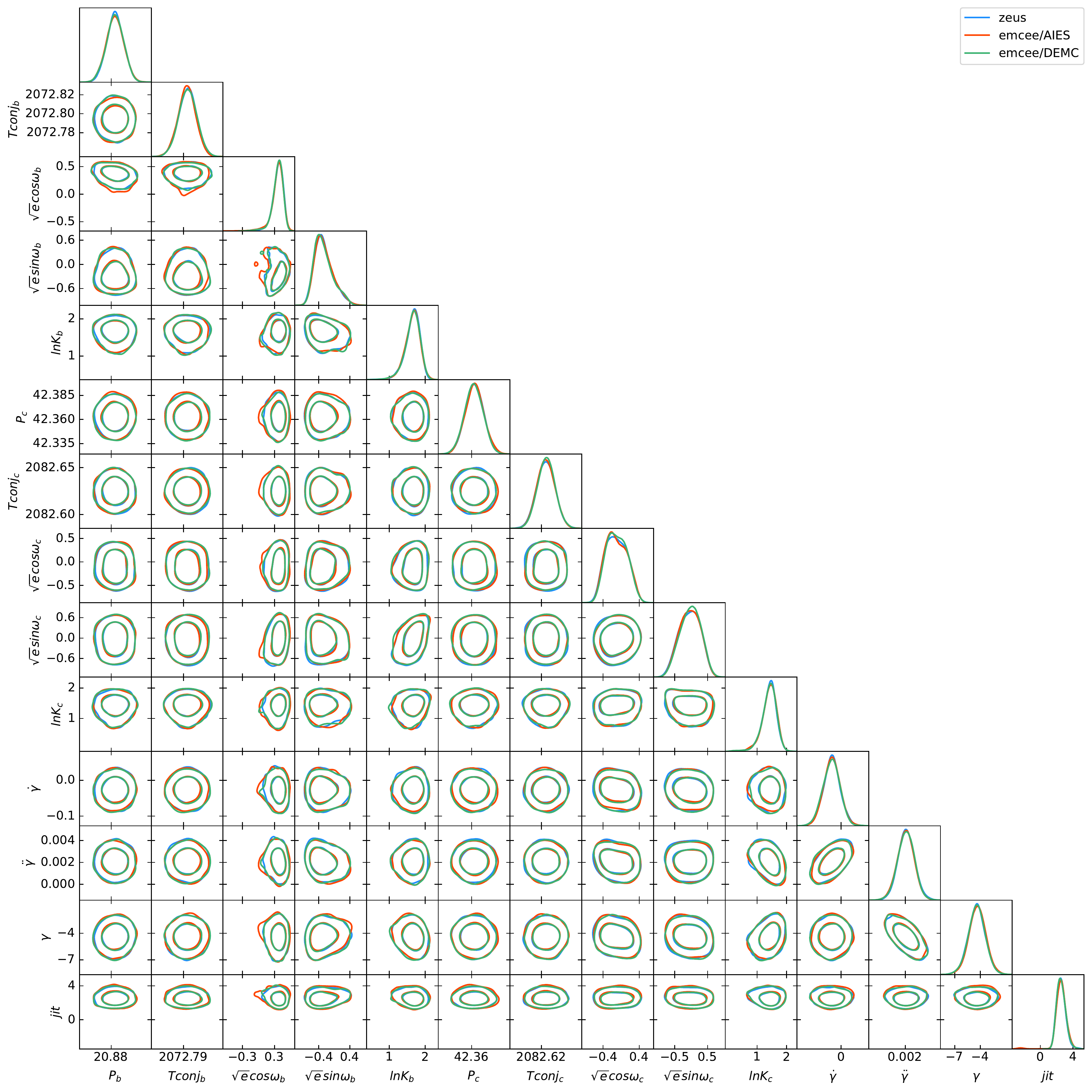}}
    \caption{A corner plot showing the 1-D and 2-D marginalised posteriors for the 14-parameter radial velocity model as produced by the three different ensemble MCMC methods.}
    \label{fig:exo}
\end{figure*}

The above discussion allows us to clearly state a crucial distinction between the three methods, which is their response to the curse of dimensionality. As the number of dimensions increases, the probability mass of a distribution is concentrated into a thin shell within the tails of the distribution (i.e. the typical set). To account for this and maintain its efficiency, a sampling method has to adjust its proposal scale -- otherwise the proposals will not be located in the typical set and thus they will not be accepted. The three methods that we mentioned so far deal with this in different ways. \texttt{emcee}/AIES's proposal scale is not adjusted and thus its proposals become increasingly inefficient in high dimensions. \texttt{emcee}/DEMC's proposal scale is adjusted based on the theoretical expectation for the case of the normal target distribution. Although both \texttt{emcee} methods perform well in this example, their sub-optimal scaling will degrade their performance in non-Gaussian target distributions as we will demonstrate in the next toy example. Finally, \texttt{zeus}'s proposal scale is continuously adapted, as the slice expands and contracts in every iteration, thus guaranteeing optimal scaling. \citet{huijser2017properties} found that the suboptimal scaling of \texttt{emcee}/AIES with the number of dimensions can introduce biases into the expectation values derived from the chains in high dimensions that are hard to diagnose. The locally adaptive nature of \texttt{zeus} allows it to avoid this problem by adjusting its proposals accordingly.

Another kind of analysis we can perform is to use the highly correlated 25--dimensional normal distribution as the target distribution and estimate the convergence rate of the three samplers. Although simple, the normal distribution is a valid approximation of many realistic astronomical posterior distributions and as such we expect the results presented in this paragraph to be applicable to a wide range of other distributions that resemble the normal distribution to some extent. We acknowledge however that the \emph{no free lunch} theorem also applies to this case, and there are bound to be cases in which the results would be qualitatively different. That being said, we initialised the walkers from a compact normal distribution (i.e. standard deviation equal to $10^{-4}$ times that of the target distribution) centred around a point along the first axis of the parameter space at a distance of $100$ standard deviations from the mode. We then measured the number of model evaluations required until the samplers have converged to the target distribution. The results for varying number of walkers are presented in Figure \ref{fig:convergence}.

In general, walkers move along directions defined by the walkers of the complementary ensemble. Thus, increasing the number of walkers offers a wider variety of available directions along which the walkers of \texttt{zeus} or \texttt{emcee} can move via slice sampling or Metropolis updates respectively. This is demonstrated in the left panel of Figure \ref{fig:convergence} in which the computational cost until convergence (i.e. number of model evaluations) for a single walker diminishes and then reaches a plateau as the number of walkers is increased. We notice however that, at the level of a single walker, the computational cost of \texttt{emcee}/AIES is significantly higher compared to that of either \texttt{zeus} or \texttt{emcee}/DEMC. This is due to the way that different samplers choose the directions along which walkers move. In particular, both \texttt{zeus} and \texttt{emcee}/DEMC define a direction vector as the difference between two walkers from the complementary ensemble, thus two walkers are required to define a direction. On the other hand, \texttt{emcee}/AIES requires only a single walker from the complementary ensemble as the direction is defined by the difference between the updated walker and the complementary one. This stark contrast between the way those samplers choose their direction vectors lies at the heart of the difference in the computational cost of \texttt{emcee}/AIES as compared to \texttt{zeus} and \texttt{emcee}/DEMC in the limit of low number of walkers. In order to dive a little deeper into this, we can compute the exact number of possible directions for all three methods. Since \texttt{emcee}/AIES requires only a single walker from the complementary ensemble the number of available directions is equal to the size of the complementary ensemble. On the other hand, \texttt{zeus}'s and \texttt{emcee}/DEMC's requirement for a pair of walkers means that the number of available directions is equal to $\binom{n}{2}$, meaning the 2--combination from a set of $n$ walkers that comprise the complementary ensemble. Clearly, as shown in Figure \ref{fig:proposals}, the latter increases faster with the size of the complementary ensemble, thus explaining the larger variety of possible directions available in the case of \texttt{zeus} and \texttt{emcee}/DEMC compared to \texttt{emcee}/AIES.

The discussion so far was about the computational cost of convergence in terms of the number of model evaluations for a single walker. Of course, the ensemble of walkers consists by definition of more than a single walker. Therefore, in order to compute the total number of model evaluations required until the ensemble converges we need to multiply the results of the single walker with the total number of walkers. Those results are presented in the right panel of Figure \ref{fig:convergence}. From this plot we can see that both \texttt{zeus} and \texttt{emcee}/DEMC converge faster when the number of walkers is close to its minimum value i.e. $2\times D$. \texttt{emcee}/AIES on the other hand prefers a higher number of walkers (i.e. $32\times D$) in order to overcome the sparsity of available directions in the limit of low number of walkers. This, however, means that even if we choose the optimal number of walkers for \texttt{emcee}/AIES it would still converge slower than either \texttt{zeus} or \texttt{emcee}/DEMC. Furthermore, we cannot know \emph{a priori} the optimal number of walkers for \texttt{emcee}/AIES unlike for \texttt{zeus} and \texttt{emcee}/DEMC in which the optimal size of the ensemble is close to $2\times D$. Finally, the faster convergence of \texttt{zeus} compared to \texttt{emcee}/DEMC can be attributed to the local adaptation that the former performs by extending the length of the slice and thus allowing larger steps in parameter space.

\subsubsection{The ring distribution}

The ring distribution defined as
\begin{equation}
\label{eq:ring}
    \ln P (x) = - \Bigg[ \frac{(x_{n}^{2} + x_{1}^{2} - a)^{2}}{b}\Bigg]^{2}  -\sum_{i=1}^{n-1} \Bigg[ \frac{(x_{i}^{2} + x_{i+1}^{2} - a)^{2}}{b}\Bigg]^{2} ,
\end{equation}
where $a=2$, $b=1$ and $n$ is the total number of parameters; this is an artificial target distribution that exhibits strong non-linear correlations between its parameters. This aspect of the ring distribution allows us to demonstrate the locally adaptive nature of \texttt{zeus}. Whereas \texttt{emcee}/AIES and \texttt{emcee}/DEMC use a single global proposal scale for all regions of the parameter space, \texttt{zeus} has the ability to adjust its proposal scale locally by expanding the slice appropriately. As expected, this will allow \texttt{zeus} to sample efficiently even in cases in which strong non-linear correlations are present. Looking at Figure \ref{fig:ring} one can see that \texttt{zeus} manages to generate multiple samples efficiently even in high dimensions. On the other hand, \texttt{emcee}/AIES and \texttt{emcee}/DEMC do not efficiently produce valid proposals: for \texttt{emcee}/AIES this leads to an inefficient random walk, characterised by small steps; for \texttt{emcee}/DEMC the acceptance rate almost vanishes beyond $D=2$. The expected squared jump distance of each method for the case of $D=25$ is shown in Table \ref{tab:table1}. It is important to note here that out of the three samplers only \texttt{zeus} manages to converge in all three cases (i.e. in 2, 10 and 25 dimensions). \texttt{emcee}/AIES and \texttt{emcee}/DEMC on the other hand converge successfully only in 2 dimensions. 

To explain this result one only has to look at the distribution of walker steps of the different methods at Figure \ref{fig:ring_sep}. \texttt{zeus}'s steps extend to large distances in parameter space whereas most of \texttt{emcee}/AIES's and \texttt{emcee}/DEMC's steps are rejected (i.e. shown as zero in the histogram). We can see that \texttt{emcee}/DEMC manages to perform some long distance steps but those are few and there is almost nothing in between. It is clear from this and the previous toy examples that the $\gamma = 2.38/\sqrt{D}$ scaling of \texttt{emcee}/DEMC's scale factor does not generalise well beyond the Gaussian case.

\subsubsection{The two-component Gaussian mixture distribution}

One other important aspect of astronomical posterior distributions is the fact that many of them exhibit multiple peaks. Multimodality can arise either from non-linear models or sparse and uninformative data. In either case, multimodal target distributions present a formidable challenge for most MCMC methods. Perhaps the simplest example of such a distribution is the two-component Gaussian mixture. In this example we will position the two, equal-mass, components at $\mathbf{-0.5}$ and $\mathbf{+0.5}$ respectively with standard deviation of $0.1$. Sampling from multimodal distributions requires two types of proposals, local proposals that sample different modes individually and global proposals that transfer walkers from one mode to the other. For this reason we will make use of \texttt{zeus}'s \texttt{GlobalMove} that uses a Dirichlet Process Gaussian Mixture model of the ensemble to efficiently propose between-mode and within-mode steps.

As seen in Figure \ref{fig:bimodal}, \texttt{zeus}'s walkers manage to move from one mode to the other frequently enough for mixing to be efficient even in the $D=25$ case. Out of \texttt{emcee}/AIES and \texttt{emcee}/DEMC, only the latter proposes valid steps from one mode to the other in the $D=2$ case. As for the $D=25$ case, one can see in Figure \ref{fig:bimodal_sep} that \texttt{zeus}'s walkers perform numerous jumps whereas \texttt{emcee}'s walkers are unable to do so. The ability of the walkers to jump from mode to mode is of paramount importance if we want to sample correctly from the target distribution. Lack of such proposals will lead to an improper probability mass ratio between the two modes and thus biased inference. The expected squared jump distance of each method for the case of $D=25$ is shown in Table \ref{tab:table1}.

Clustering-based proposals have also been applied to MH-type ensemble MCMC methods but as shown in \citet{karamanis2020ensemble}, they fail to generate valid proposals in problems with moderate number of dimensions. The reason is, as discussed in Section \ref{sec:tests}, that MH has to propose a valid point in the other mode. In other words, whereas Ensemble Slice Sampling only needs to determine the direction of the other mode relative to the chosen walker correctly, MH needs to guess both the direction and the distance, a task that rapidly becomes very hard as the number of dimensions rises.

\subsubsection{The Student's $t$-distribution}

The fourth toy example tests the case in which the target distribution is characterised by heavy-tails. In order to demonstrate \texttt{zeus}'s ability to sample efficiency is such cases we chose to use the multivariate Student's $t$-distribution with $2$ degrees of freedom. The aforementioned density exhibits heavier tails than a normal distribution which means that it is more likely to produce samples that are far away from the mean. The $t$-distribution arises when estimating the mean of a normally distributed sample with unknown standard deviation and small size. The probability density function of a $p$--dimensional Student's $t$-distribution with $\nu$ degrees of freedom is given by:
\begin{equation}
    \label{eq:student}
    P(x) = \frac{\Gamma[(\nu+p)/2]}{\Gamma(\nu/2)\nu^{p/2}\pi^{p/2}|\boldsymbol{\Sigma}|^{1/2}}\exp\bigg[ 1 + \frac{1}{\nu}(\mathbf{x}-\boldsymbol{\mu})^{T}\boldsymbol{\Sigma}^{-1}(\mathbf{x}-\boldsymbol{\mu})\bigg]^{-\frac{(\nu+p)}{2}}\,,
\end{equation}
where $\boldsymbol{\Sigma}$ is the $p\times p$ positive semi-definite shape matrix and $\boldsymbol{\mu}$ is the mean vector.

We sampled the above distribution using the three samplers in $2$,  $10$ and $25$ dimensions respectively as shown in Figure \ref{fig:student}. The diagonal elements of shape matrix $\boldsymbol{\Sigma}$ were set to $1$ and the off-diagonal elements to $0.95$. The mean vector $\boldsymbol{\mu}$ was set to $\mathbf{0}$. All three samplers managed to sample efficiently in $2$,  $10$ and $25$ dimensions as shown in Figure~\ref{fig:student} and Table~\ref{tab:table1}. Overall, \texttt{zeus} was the most efficient method with \texttt{emcee}/AIES being second and \texttt{emcee}/DEMC last. One can see from Figure \ref{fig:student_sep} that the distributions of steps of \texttt{zeus} and \texttt{emcee}/AIES are very similar whereas that of \texttt{emcee}/DEMC is substantially shorter. Unlike the previous toy examples in which the proposal strategy of \texttt{emcee}/AIES was causing it to overshoot the bulk of posterior mass, in the case of the heavy-tailed $t$-distribution more proposals are accepted. On the other hand, \texttt{emcee}/DEMC's proposals which are optimised for Gaussian targets are more conservative in the case of the $t$-distribution and they do not extend far away. As also demonstrated in the previous toy examples, the locally adaptive nature of \texttt{zeus} allows it to perform efficient proposals that span large distances in parameter space.

\subsubsection{The truncated normal distribution}

The fifth and final toy example tests the case in which the target distribution is bounded from below or above. We chose to employ a truncated normal distribution similar to the one used in the first toy example, with the additional constraint being that $x > 0$. This effectively introduces a hard boundary along all dimensions. One of the reasons that we study this distribution is to assess the bias introduced by the presence of the hard boundary.

We sampled the above distribution using the three samplers in $2$,  $10$ and $25$ dimensions respectively as shown in Figure \ref{fig:truncated}. The diagonal elements of the covariance matrix were set to $1$ and the off-diagonal to $0.95$. The mean vector $\boldsymbol{\mu}$ was set to $\mathbf{0}$. All three samplers managed to sample efficiently in $2$,  $10$ and $25$ dimensions as shown in Figure~\ref{fig:truncated} and Table~\ref{tab:table1}.  Overall, \texttt{zeus} was the most efficient method with \texttt{emcee}/DEMC being second and \texttt{emcee}/AIES last. One can see from Figure \ref{fig:truncated_sep} that the distributions of steps of \texttt{zeus} and \texttt{emcee}/AIES are very similar whereas that of \texttt{emcee}/AIES is slightly shorter. As shown in the right panels of Figure \ref{fig:truncated} \texttt{zeus} exhibits the least amount of bias compared to \texttt{emcee}/AIES and \texttt{emcee}/DEMC. In practical astronomical examples however, only one or two parameters would usually be bounded (e.g. the neutrino mass in galaxy clustering analyses) and thus unbiased sampling would be easier to perform by either of the three samplers.

\subsection{Real astronomical analyses}

The previous section employs toy examples in order to exhibit various scenarios that might emerge during sampling, and shows how \texttt{zeus} is better equipped to handle them. To demonstrate the efficiency of \texttt{zeus} compared to other samplers in realistic target distributions, we chose two common astronomical inference problems as the testing ground. Those are the cases of baryon acoustic oscillation (BAO) parameter inference and exoplanet parameter estimation.

We used the same three samplers in our comparison, namely \texttt{emcee} with AIES and DEMC, and of course \texttt{zeus}. We performed three distinct tests:

\begin{itemize}
    \item The first test was to estimate the efficiency for each sampler, defined as the number of independent samples produced per log-likelihood evaluation. To this end, we ran the MCMC procedure 5 times for each sampler and computed the mean efficiency using the estimated autocorrelation time of the chains. The autocorrelation time was estimated using the method presented in \citet{karamanis2020ensemble}.
    \item The second test relates to the convergence rate of the three algorithms. As a measure of convergence rate, we adopt the inverse of the number of iterations required until all the convergence criteria specified below are met. In order to estimate the mean convergence rate we ran the sampling procedure 40 times for each sampler initialising the walkers close to the \textit{Maximum a Posteriori} (MAP) estimate. 
    \item Finally, we tested the sensitivity of the samplers to the initial conditions by running 40 realisations with the walkers initialised from a small sphere (of radius $10^{-4}$) around a randomly chosen point in the prior volume, counting how many of those attempts led to converged chains before a predetermined number of likelihood evaluations.
\end{itemize}

To determine whether a chain has converged we used four different metrics: the Gelman-Rubin split-$R$ statistic \citep{gelman1992inference, gelman2013bayesian} using four independent ensembles of walkers; the Geweke test \citep{geweke1992evaluating}; a minimum length of the chain as a multiple of the integrated autocorrelation time (IAT); as well as an upper bound on the rate of change of the IAT. Only the second half of the chains was used to evaluate the aforementioned criteria. The number of walkers used in both examples was close to the minimum value of $2\times D$ as specified below. As we will discuss in Section \ref{sec:discussion} this often leads to faster convergence.

\subsubsection{Cosmological inference}

The particular inference problem that we consider here is that of the anisotropic BAO parameter inference using estimates of the galaxy power spectrum. The data we used come from the 12th data release (DR12) of the high-redshift North Galactic Cap (NGC) sample as observed by the Sloan Digital Sky Survey (SDSS) \citep{SDSSIII} Baryon Oscillation Spectroscopic Survey (BOSS) \citep{BOSS}. Our analysis follows closely that of \citet{beutler2017clustering} with the difference that we chose not to fix any parameters and fit the hexadecapole multipole of the power spectrum as well as the monopole and quadrupole. Those choices were made solely to render the problem more challenging. Indeed the inclusion of the hexadecapole does not contribute any additional constraining power for the data that we used. However, such extended models will prove useful when analysing data from larger galaxy surveys such as DESI \citep{DESI}. In terms of Bayesian inference, the problem has 22 free parameters. The results of our analysis are consistent with those of \citet{beutler2017clustering}. We used weakly informative flat (uniform) priors for all parameters except for the two scaling parameters, $\alpha_{\parallel}$ and $\alpha_{\bot}$ for which we used normal (Gaussian) priors. We used $50$ walkers in total.

In terms of efficiency, \texttt{zeus} generates at least 5 independent samples for each one generated by \texttt{emcee}/DEMC and at least 9 for each one generated by \texttt{emcee}/AIES factoring in the different computational cost of the methods. As for the convergence rate, \texttt{zeus} converges more than 3 times faster than either \texttt{emcee} variant. Finally, we found that \texttt{zeus} is less sensitive to the initialisation than either of the other two methods. In particular, out of the 40 tests conducted with different initialisation, \texttt{zeus} converged 36 times, \texttt{emcee}/DEMC 14 times and \texttt{emcee}/AIES 7 times prior to the predetermined maximum number of likelihood evaluations (i.e. $5\times 10^6$ in this case). The aforementioned results are presented in detail in Table \ref{tab:table2}. The 1-D and 2-D marginal posterior distributions are shown in Figure \ref{fig:bao} demonstrating the agreement between the three methods\footnote{No upper limit on the number of likelihood evaluations or iterations was used for this run and convergence was diagnosed using all the metrics that we introduced.}.

\begin{table}
    \centering
    \caption{The table shows a comparison of \texttt{emcee}/AIES, \texttt{emcee}/DEMC and \texttt{zeus} in terms of the inverse efficiency (i.e. reciprocal of the number of independent samples per model evaluation or the autocorrelation time estimate times the average number of model evaluations per iteration per walker), the convergence cost (i.e. number of model evaluations until convergence) and the convergence fraction (i.e. fraction of converged chains for given maximum number of model evaluations).}
    \def\arraystretch{1.1}
    \begin{tabular}{lcccc}
        \toprule[0.75pt]
         & \texttt{emcee}/AIES   & \texttt{emcee}/DEMC   & \textbf{zeus}  \\
        \midrule[0.5pt]
        \multicolumn{4}{l}{Cosmological inference} \\
        \midrule[0.5pt]
        efficiency$^{-1}$    &    12140    &    6750    &   $\mathbf{1320}$   \\
        convergence cost & $24\times10^{5}$ & $22\times10^{5}$ & $\mathbf{6.6\times10^{5}}$   \\
        convergence fraction   &    7/40   &    14/40    & $\mathbf{36/40}$  \\
        \midrule[0.5pt]
        \multicolumn{4}{l}{Exoplanet inference} \\
        \midrule[0.5pt]
        efficiency$^{-1}$    &    $1386$    &    $338$    &   $\mathbf{47}$   \\
        convergence cost & $36.0\times 10^{2}$ & $17.1\times 10^{2}$ & $\mathbf{4.8\times 10^{2}}$   \\
        convergence fraction   &    23/40    &    29/40    &  $\mathbf{38/40}$ \\
        \bottomrule[0.75pt]
        \end{tabular}
    \label{tab:table2}
\end{table}

\subsubsection{Exoplanet inference}

Another common application of MCMC methods in astronomy is the problem of exoplanet parameter inference through modelling of Keplerian orbits and radial velocity time series data. In this section we demonstrate the performance of \texttt{zeus} using a two-planet model with $14$ free parameters and real data from the K2-24 (EPIC-203771098) extrasolar system \citep{k224} that is known to host two exoplanets. We used the popular \texttt{Python} package \texttt{RadVel} \citep{radvel} for the Keplerian modelling of the planetary orbits. The results of our analysis are consistent with published constraints for the aforementioned extrasolar system \citep{k224}. We used $30$ walkers in total for sampling.

We performed the same suite of tests as in the cosmological inference case. In terms of efficiency, \texttt{zeus} generates more than $7$ independent samples per each one generated by \texttt{emcee}/DEMC and more than $29$ independent samples per each one generated by \texttt{emcee}/AIES. As for the convergence rate, \texttt{zeus} converges $7.5$ times faster than \texttt{emcee}/AIES and $3.5$ faster than \texttt{emcee}/DEMC on average. Finally, we found again that \texttt{zeus} is less sensitive to the specific initialisation of the walkers. In particular, out of the 40 tests conducted with different initialisation, \texttt{zeus} converged 38 times, \texttt{emcee}/DEMC 29 times and \texttt{emcee}/AIES 23 times prior to the predetermined maximum number of likelihood evaluations (i.e. $5\times 10^3$ in this case). Detailed results about the values of the used metrics are shown in Table \ref{tab:table2}. The 1-D and 2-D marginal posterior distributions are shown in Figure \ref{fig:exo}, demonstrating the agreement between the three methods.

\section{Discussion}
\label{sec:discussion}
Following the analysis we conducted in Section \ref{sec:tests} using the normal distribution there are two important questions that need to be answered about the initialisation of the walkers. First, how many walkers are necessary and, second, how to choose the initial positions of the walkers. Although there are many ways of answering those questions and there is no consistent solution that works for all target distributions, we will try to provide some general rules and heuristics to help ease the task of choosing the number and initial positions of the walkers for most cases.

Let us first discuss the effect of the number of the walkers on the general performance of \texttt{zeus}. Naively, one might expect that the minimum number of walkers should be $D+1$, where $D$ is the number of dimensions. However, the ensemble splitting technique, which was introduced in Section \ref{sec:tests} to render the algorithm parallelisable, requires at least $2\times D$ walkers in order to produce $2$ linearly independent samples. If a smaller number is chosen then the walkers can be trapped in a lower--dimensional hyper--plane of the parameter space, being unable to sample properly and leading to erroneous results. Although there is no upper bound on the number of walkers, we recommend to use between two to four times the number of dimensions. The reason is that increasing the number of dimensions can increase the cost of the burn-in period as we explained in detail in Section \ref{sec:tests}. Ideally, one wants to use the minimum number (or close to that) of walkers until the burn-in period is over and then increase the number of walkers to rapidly produce a great number of independent samples. It is also worth noting that in cases in which either non-linear correlations or multiple modes are present it is recommended to use more walkers (e.g. 4-8 times the number of parameters for a bimodal target distribution).

As for the initialisation of the walkers, there are many ways to choose their starting positions ranging from prior sampling to more localised initial positions. Empirical tests indicate that the latter often outperforms the former (i.e. leads to shorter burn-in periods). That is not surprising since the total probability of a prior-sampled initialisation can be very small when the number of parameters is high. In particular we found that initialising the walkers from a tight region in parameter space (i.e. normal distribution with small variance) consistently leads to good performance. For low to moderate dimensional problems initialising the walkers from a tight ball around the \textit{Maximum A Posteriori} (MAP) estimate can substantially reduce the burn-in period~\citep{foreman2013emcee}.

Finally, while \texttt{emcee}/AIES and \texttt{emcee}/DEMC can sample acceptably from most target distributions with $D\lesssim 20$, the efficient scaling of \texttt{zeus} with the number of parameters allows us to extend this range and efficiently test more complicated models \citep{karamanis2020ensemble}. Like most gradient-free methods, \texttt{zeus} will fail to sample efficiently in very high dimensional problems in which $D=\mathcal{O}(10^{2})$. In such cases, more sophisticated algorithms (e.g. tempering, block updating, Hamiltonian dynamics etc.) need to be used \citep{2018arXiv180402719R}.

\section{Conclusions}
\label{sec:conclusions}

The aim of this project was to develop a tool that could facilitate Bayesian parameter inference in computationally demanding astronomical analyses and tackle the challenges posed by the complexity of the models and data that are often used by astronomers. To this end, we introduced \texttt{zeus}, a parallel, general-purpose and gradient-free \texttt{Python} implementation of Ensemble Slice Sampling.

After introducing the method in Section \ref{sec:ess}, we thoroughly demonstrated its performance compared to two popular alternatives (i.e. \texttt{emcee} with affine-invariant ensemble sampling and differential evolution Metropolis) using a variety of artificial and realistic target distributions in Section \ref{sec:tests}. The artificial toy examples helped to shed light on the general behaviour of the samplers in target distributions characterised by linear and non-linear correlations as well as multimodal densities. When compared to \texttt{emcee}/AIES and \texttt{emcee}/DEMC in the problems of Baryon Acoustic Oscillation parameter inference and exoplanet radial velocity fitting, \texttt{zeus} consistently converges faster (i.e. its burn-in is shorter by a factor of at least 3), it is less sensitive to the initialisation of the walkers and generates substantially more independent samples per likelihood evaluation (i.e. approximately $\times 9$ and  $\times 29$ speed-up compared to \texttt{emcee}/AIES in the cosmological and exoplanet examples, respectively). 

We have shown that \texttt{zeus} performs similarly or better than existing MCMC methods in a range of problems. We hope that \texttt{zeus} will prove useful to the astronomical and cosmological community by complementing existing approaches and facilitating the study of novel models and data over the coming years. \texttt{zeus} is publicly available at \url{https://github.com/minaskar/zeus} with detailed documentation and examples that can be found at \url{https://zeus-mcmc.readthedocs.io}.

\section*{Acknowledgements}

The authors would like to express their gratitude to anyone who contributed to the early success of \texttt{zeus} by incorporating it into their own research. In particular, we thank Andrius Tamosiunas, Jamie Donald-McCann, Mike Wang and Obinna Umeh for providing feedback on an early version of the code. We also extend our gratitude to Johannes Buchner for providing constructive comments and suggestions. This project has received funding from the European Research Council (ERC) under the European Union's Horizon 2020 research and innovation programme (grant agreement 853291). FB is a University Research Fellow.

\texttt{zeus} and this work have benefited from a variety of \texttt{Python} packages including \texttt{numpy} \citep{van2011numpy}, \texttt{scipy} \citep{virtanen2020scipy}, \texttt{matplotlib} \citep{hunter2007matplotlib}, \texttt{seaborn} \citep{Waskom2021seaborn}, \texttt{getdist} \citep{lewis2019getdist}, \texttt{sklearn} \citep{pedregosa2011scikit}, \texttt{tqdm} \citep{da2019tqdm}, and \texttt{mpi4py} \citep{dalcin2011parallel}.

\section*{Data Availability}

All data used in this work are publicly available. Power spectrum estimates, covariance matrices and window functions used in the cosmological inference example are available at \url{http://www.sdss3.org/science/boss_publications.php}. Radial velocity measurements of the K2-24 system used in the exoplanet inference are available as part of the \texttt{RadVel} package at \url{https://github.com/California-Planet-Search/radvel}.




\bibliographystyle{mnras}
\bibliography{references} 




\appendix

\section{\texttt{cronus}}

Along with \texttt{zeus}, we introduce \texttt{cronus}, a \texttt{Python} package that wraps \texttt{zeus} as well as other samplers (e.g. \texttt{emcee}). \texttt{cronus} is designed to be used via the terminal through simple, human-readable parameters files. \texttt{cronus} implements an automated suite of convergence diagnostics and all its methods are massively parallel using the Message Passing Interface (MPI). Our aim is to automate parameter inference and facilitate highly reproducible analyses for current and future astronomical surveys such as the Dark Energy Spectroscopic Instrument (DESI) \citep{DESI} \texttt{cronus} is available at \url{https://cronus-mcmc.readthedocs.io}.


\bsp	
\label{lastpage}
\end{document}